\documentclass[a4paper,11pt]{revtex4}
\usepackage{graphicx} 
\usepackage{amsmath} 
\usepackage{amssymb} 
\usepackage{bm} 
\usepackage{dcolumn}
\usepackage{color}
\usepackage{mathrsfs}
\usepackage{amsfonts}
\usepackage{varioref}
\usepackage{mathrsfs}
\usepackage{graphicx}
\usepackage{latexsym}
\usepackage{amsmath}
\usepackage{amssymb}
\usepackage{textcomp}
\usepackage{amsbsy}
\usepackage{graphics}
\usepackage{epstopdf}
\usepackage{color}
\usepackage[caption=false]{subfig}

\RequirePackage[colorlinks,citecolor=blue,urlcolor=magenta,linkcolor=blue]{hyperref}
\input epsf

\allowdisplaybreaks[4]
\begin{document}

\tolerance=5000

\title{Second law of horizon thermodynamics during cosmic evolution}

\author{Sergei~D.~Odintsov$^{1,2}$\,\thanks{odintsov@ieec.uab.es},
Tanmoy~Paul$^{3}$\,\thanks{tanmoy.paul@visva-bharati.ac.in},
Soumitra~SenGupta$^{4}$\,\thanks{tpssg@iacs.res.in}} \affiliation{
$^{1)}$ ICREA, Passeig Luis Companys, 23, 08010 Barcelona, Spain\\
$^{2)}$ Institute of Space Sciences (ICE, CSIC) C. Can Magrans s/n, 08193 Barcelona, Spain\\
$^{3)}$ Department of Physics, Visva-Bharati University, Santiniketan 731235\\
$^{4)}$ School of Physical Sciences, Indian Association for the Cultivation of Science, Kolkata-700032, India}


\tolerance=5000

\begin{abstract}
We examine the second law of thermodynamics in the context of horizon cosmology, in particular, whether the change of total entropy (i.e. the sum of the entropy for the apparent horizon and the entropy for the matter fields) proves to be positive with the cosmic expansion of the universe. The matter fields inside  the horizon obey the thermodynamics of an open system as the matter fields has a flux through the apparent horizon, which is either outward or inward depending on the background cosmological dynamics. Regarding the entropy of the apparent horizon, we consider different forms of the horizon entropy like the Tsallis entropy, the R\'{e}nyi entropy, the Kaniadakis entropy, or even the 4-parameter generalized entropy; and determine the appropriate conditions on the respective entropic parameters coming from the second law of horizon thermodynamics. The constraints on the entropic parameters are found in such a way that it validates the second law of thermodynamics during a wide range of cosmic era of the universe, particularly from inflation to radiation dominated epoch followed by a reheating stage. Importantly, the present work provides a model independent way to constrain the entropic parameters directly from the second law of thermodynamics for the apparent horizon.
\end{abstract}

\maketitle

\section{Introduction}\label{SecI}

One of the the distinctive features of Bekenstein-Hawking entropy of a black hole is that it depends on the area of the event horizon \cite{Bekenstein:1973ur,Hawking:1975vcx,Bardeen:1973gs,Wald:1999vt}, unlike  the classical thermodynamics where the entropy of a thermodynamic system depends on the volume of the same under consideration. Based on such an interesting feature of  Bekenstein-Hawking entropy, and depending on the non-additive statistics, various other form of entropies  have been proposed  such as Tsallis \cite{Tsallis:1987eu} and the R\'{e}nyi \cite{Renyi} entropies. The Barrow entropy has been recently proposed in \cite{Barrow:2020tzx} to capture the fractal nature of a black hole originated from the quantum gravitational effects. Furthermore, the Sharma-Mittal (which is essentially a combined form of the Tsallis and the R\'{e}nyi entropies) \cite{SayahianJahromi:2018irq}, the Kaniadakis entropy \cite{Kaniadakis:2005zk}, the entropy in the context of Loop Quantum Gravity (LQG) \cite{Majhi:2017zao} are some other well known description of entropies which are some functions of Bekenstein-Hawking entropy variable. Despite their different forms, all these entropies share some common properties, like --- (a) they converge to the Bekenstein-Hawking entropy for some suitable limit of the respective entropic parameters, and (b) they are monotonically increasing function with respect to the Bekenstein-Hawking variable. Such common properties immediately lead to a natural question that whether there exists any generalized entropy which can generalize all these known entropies proposed so far. In this route, few parameters dependent generalized entropies have been proposed in \cite{Nojiri:2022aof,Nojiri:2022dkr,Odintsov:2022qnn}, which is a generalized form of all the aforementioned entropies  for appropriate limits of the entropic parameters. However according to the conjecture made in \cite{Nojiri:2022dkr}, a 4-parameter dependent entropy is the minimal version of generalized entropy. Some possible implications of generalized entropies to cosmology as well as to black hole physics are discussed in \cite{Nojiri:2022dkr,Odintsov:2022qnn,Nojiri:2022nmu,Odintsov:2023qfj,Odintsov:2023vpj}.

In the context of cosmology, the homogeneous and isotropic universe acquires an apparent horizon which, being a null surface, divides the observable universe from the unobservable one.
Thus in  analogy of black hole thermodynamics, the apparent horizon in cosmology may also be associated with an entropy \cite{Cai:2005ra,Akbar:2006kj,Cai:2006rs,Paranjape:2006ca,Jamil:2009eb,Cai:2009ph,Jamil:2010di,Gim:2014nba,DAgostino:2019wko,Nojiri:2023wzz,Sanchez:2022xfh,Cognola:2005de}. Furthermore the entropic cosmology proves to be equivalent to holographic cosmology with suitable holographic cut-offs which actually depends on the entropy function under consideration \cite{Nojiri:2021iko}. In this regard holographic cosmology, initiated by Witten and Susskind in \cite{Witten:1998qj,Susskind:1998dq,Fischler:1998st}, earned a lot of attention as it is directly related to the entropy construction. The most intriguing question in modern
cosmology is to explain the accelerating phases of the universe during two extreme curvature regimes, namely the
inflation and the dark energy era of the universe. The holographic cosmology sourced from the aforementioned
entropies successfully explain the dark energy era of the universe for constant as well as for variable exponents of
the entropy functions, and generally known as the holographic dark energy (HDE) model \cite{Li:2004rb,Wang:2016och,Pavon:2005yx,Nojiri:2005pu,Malekjani:2012bw,Khurshudyan:2016gmb,Gao:2007ep,Zhang:2005hs,Li:2009bn,Feng:2007wn,Zhang:2009un,Lu:2009iv,Nojiri:2017opc}. Besides the dark energy era, the holographic cosmology also turns out to be a suitable candidate to explain inflationary era during the early stage of the universe when the size of the universe was  small and the holographic energy density is good enough to trigger an inflation of the universe \cite{Nojiri:2019kkp,Oliveros:2019rnq}. More interestingly, the holographic cosmology provides  unification of an early inflation to a late dark energy era of the universe in a covariant manner \cite{Nojiri:2020wmh}. All these works reflect the intense interest on holographic or equivalently on
entropic cosmology corresponding to various entropy functions.

In the arena of entropic cosmology, the cosmological field equations are based on the first law of thermodynamics of the apparent horizon. However a consistent cosmological scenario, also demands the validation of the second law of horizon thermodynamics, i.e., whether the change of total entropy (which is the sum of the horizon entropy and the entropy of the matter fields) proves to be positive with the cosmic expansion of the universe. In the present paper we intend to do this. In this regard, the matter fields inside the horizon obey the thermodynamics of an open system as the matter fields can have an inward or outward  a flux through the apparent horizon. Regarding the entropy for the apparent horizon, we will consider different forms of the horizon entropy like the Tsallis entropy, the R\'{e}nyi entropy, the Kaniadakis entropy, or even the 4-parameter generalized entropy; and will determine the appropriate conditions on the respective entropic parameters coming from the second law of horizon thermodynamics. The constraints on the entropic parameters are found in such a way that it validates the second law of thermodynamics during a wide range of cosmic evolution of the universe, particularly from $\mathrm{inflation}~\rightarrow~\mathrm{reheating}~\rightarrow~\mathrm{radiation~era}$ respectively.

The paper is organized as follows: in Sec.~[\ref{SecII}], we will discuss the basic formalism of apparent horizon thermodynamics and will determine the cosmological field equations corresponding to a general form of horizon entropy. Sec.~[\ref{SecIII}] is reserved for the thermodynamics of the matter fields inside the horizon. In Sec.~[\ref{sec-total-change}], we will focus on the total entropy and its change with the cosmic time. The paper will end with some concluding remarks in Sec.~[\ref{sec-conclusion}].

\section{Thermodynamics of apparent horizon and cosmological field equations}\label{SecII}

We consider the $(3+1)$ dimensional spatially flat FLRW universe, whose metric is given by,
\begin{align}
\label{dS7}
ds^2 = \sum_{\mu,\nu=0,1,2,3} g_{\mu\nu} dx^\mu dx^\nu = - dt ^2 + a( t )^2 \left( d r ^2 + r ^2 {d\Omega_{2}}^2 \right) \, ,
\end{align}
where ${d\Omega_{2}}^2$ is the line element of a $2$ dimensional sphere of unit radius (particularly on the surface of the sphere).
We also define
\begin{align}
\label{dS7B}
d{s_\perp}^2 = \sum_{M,N=0,1} h_{\mu\nu} dx^M dx^N = - dt ^2 + a( t )^2 d r ^2 \, .
\end{align}
The radius of the apparent horizon $R_\mathrm{h}=R\equiv a(t)r$ for the FLRW universe is given by the solution of the equation 
$h^{MN} \partial_M R \partial_N R = 0$ (see \cite{Cai:2005ra, Akbar:2006kj}) which immediately leads to,
\begin{align}
\label{dS14A}
R_\mathrm{h}=\frac{1}{H}\, ,
\end{align}
with $H\equiv \frac{1}{a}\frac{da}{d t }$ represents the Hubble parameter of the universe. It may be noted that the apparent horizon in the case of a spatially flat FLRW universe becomes equal to the Hubble radius. The surface gravity $\kappa$ on the apparent horizon is defined as \cite{Cai:2005ra}
\begin{align}
\label{SG3}
\kappa= \left. \frac{1}{2\sqrt{-h}} \partial_M \left( \sqrt{-h} h^{MN} \partial_N R \right) \right|_{R=R_\mathrm{h}}\, .
\end{align}
For the metric of Eq.~(\ref{dS7}), we have $R=a r $ and obtain 
\begin{align}
\label{SG2}
\kappa = - \frac{1}{R_\mathrm{h}} \left\{ 1 + \dot{H}\left(\frac{{R_\mathrm{h}}^2}{2}\right) \right\} \, ,
\end{align}
where the following expression is used, 
\begin{align}
\label{dS14AB}
\dot R_\mathrm{h} = - H\dot{H} {R_\mathrm{h}}^3 \, ,
\end{align}
The surface gravity of Eq.~(\ref{SG2}) is related with the temperature via $T_\mathrm{h} = \left|\kappa\right|/(2\pi)$, i.e.,
\begin{align}
\label{AH2}
T_\mathrm{h} \equiv \frac{\left| \kappa \right|}{2\pi} 
= \frac{1}{2\pi R_\mathrm{h}} \left| 1 - \frac{\dot R_\mathrm{h}}{2 H R_\mathrm{h}} \right|
= \frac{H}{2\pi} \left| 1 + \frac{\dot{H}}{2H^2} \right|\, ,
\end{align}
in terms of the Hubble parameter and its derivative. 
Consequently, we may associate an entropy ($S_\mathrm{h}$) to the apparent horizon, which in turn follows the thermodynamic law given by \cite{Akbar:2006kj},
\begin{align}
T_\mathrm{h}dS_\mathrm{h} = -dE + WdV\, ,
\label{law-1}
\end{align}
where $V = \frac{4}{3}\pi R_\mathrm{h}^3$ is the volume of the space enclosed by the apparent horizon (for a different thermodynamic law, see \cite{Nojiri:2023wzz}). Moreover, $E = \rho V$ is the total internal energy of the matter fields inside the horizon, and $W = \frac{1}{2}\left(\rho - p\right)$ denotes the work density
by the matter fields \cite{Akbar:2006kj,Nojiri:2023wzz}. Eq.~(\ref{law-1}) argues that the horizon entropy exists due to the reasons --- (a) decrease of internal energy of the matter fields inside of the horizon, denoted by the term $-dE$ in the rhs of Eq.~(\ref{law-1}), and (b) the work done by the matter fields coming through the term $WdV$. The decrease of internal energy of the matter fields as well as the work done by the matter fields may be regarded as an energy flux through the apparent horizon. Since the apparent horizon divides the observable universe from the unobservable one, such energy flux can be thought as some information loss of the observable universe, which in turn gives rise to an entropy of the horizon. However here we would like to mention that a proper understanding of microscopic origin for the entropy of the apparent horizon still elludes us and one may see \cite{Nojiri:2023bom} for some progress in this regard.

Following points need to be mentioned regarding the temperature of the apparent horizon mentioned in Eq.~(\ref{AH2}): (a) the form of $T_\mathrm{h}$ in Eq.~(\ref{AH2}) (coming from the surface gravity of the apparent horizon) is different from that of in \cite{Gong:2007md} where the authors used $T_\mathrm{h} = H/(2\pi)$. Actually in the context of black hole thermodynamics, the gravitational field equations can be interpreted from the thermodynamics of the event horizon, in which case, the temperature of the horizon is proportional to the surface gravity of the same \cite{Bardeen:1973gs,Paranjape:2006ca,Padmanabhan:2002sha}. In this regard, the important point is that if $gravity$ has a $thermodynamic$ connection owing to the presence of a horizon, then the temperature of the horizon should have a universal definition both in black hole as well as in cosmological context. Keeping this in mind and from the analogy of black hole thermodynamics, here in the cosmological scenario, we similarly consider the temperature of the apparent horizon to be the surface gravity of the same (as per Eq.~(\ref{AH2})) which is also widely accepted in \cite{Akbar:2006kj,Cai:2006rs,Nojiri:2022nmu,Nojiri:2023wzz,Sanchez:2022xfh}. Moreover it may be noted that the $T_\mathrm{h}$ in Eq.~(\ref{AH2}) reduces to that of used in \cite{Gong:2007md} for a de-Sitter universe. (b) Eq.~(\ref{AH2}) clearly indicates that $T_\mathrm{h}$ goes to zero during the radiation era (i.e. for $H \propto a^{-2}$, $a$ is the scale factor of the universe), in which case, the trace of the energy-momentum tensor of the matter field inside of the horizon vanishes. This is analogous to the case of an extremal Reissner-Nordstrom black hole, in the context of black hole thermodynamics, where the temperature of the event horizon vanishes due to $Q=M$ (where $Q$ and $M$ represent charge and mass of the black hole respectively). Therefore the radiation dominated era may be considered as an extremal case in the sector of horizon cosmology. As a result, the rhs of Eq.~(\ref{law-1}) consequently vanishes and thus the thermodynamic law (\ref{law-1}) becomes a trivial one during the radiation era, due to which, Eq.~(\ref{law-1}) is unable to extract the change of horizon entropy (i.e. $dS_\mathrm{h}$) when the universe undergoes through radiation dominated era. (c) Finally we would like to mention that $T_\mathrm{h}$ always comes with a positive value (including $T_\mathrm{h} = 0$ for a radiation dominated universe) due to the absolute value of $\kappa$.

In the context of entropic cosmology, the thermodynamics of the apparent horizon governed by Eq.~(\ref{law-1}) fixes the gravitational field equations, and depending on the form of $S_\mathrm{h}$, the field equations get modified. However irrespective of the form, $S_\mathrm{h}$ shares some common properties like :
\begin{itemize}
 \item $S_\mathrm{h}$ is a monotonic increasing function of the Bekenstein-Hawking entropy variable $S = A/(4G)$ (where $A = 4\pi R_\mathrm{h}^2$ denotes the area of the apparent horizon),

 \item $S_\mathrm{h}$ goes to zero in the limit of $S \rightarrow 0$, which can be thought as equivalent of the third law of thermodynamics.
\end{itemize}
In the following, we derive the gravitational field equations from Eq.~(\ref{law-1}) for a general form of the horizon entropy given by $S_\mathrm{h}$. Taking $E = \rho V$ and $W = \frac{1}{2}\left(\rho - p\right)$ into account, Eq.~(\ref{law-1}) can be written by,
\begin{eqnarray}
 T_\mathrm{h}\dot{S}_\mathrm{h} = -\dot{\rho}V - \frac{1}{2}\left(\rho + p\right)\dot{V}~~,
 \label{HE-1}
\end{eqnarray}
where the overdot symbolizes $\frac{d}{dt}$ of the respective quantity. Because of the energy conservation (local conservation) of the matter fields inside of the horizon, we have $\nabla_{\mu}T^{\mu\nu} = 0$ (where $\nabla_{\mu}$ is the covariant derivative formed by the metric $g_{\mu\nu}$, and $T^{\mu\nu}$ is the energy-momentum tensor of the matter field) which, due to the FLRW metric of Eq.~(\ref{dS7}), takes the following form,
\begin{eqnarray}
 \dot{\rho} + 3H\left(\rho + p\right) = 0~~,
 \label{conservation-matter}
\end{eqnarray}
Using the above expression into Eq.~(\ref{HE-1}), one gets,
\begin{eqnarray}
 T_\mathrm{h}\dot{S}_\mathrm{h} = \left(\rho + p\right)\left\{3HV - \frac{\dot{V}}{2}\right\}~~,
 \label{HE-2}
\end{eqnarray}
which, owing to $V=\frac{4}{3}\pi R_\mathrm{h}^3$, takes the following form:
\begin{eqnarray}
 \dot{S}_\mathrm{h} = \frac{8\pi}{H^3}\left(\rho + p\right)~~.
 \label{HE-3}
\end{eqnarray}
As mentioned above that $S_\mathrm{h}$ is a function of the Bekenstein-Hawking entropy variable $S$, and thus Eq.~(\ref{HE-3}) can be expressed by,
\begin{eqnarray}
 \dot{H}\left(\frac{\partial S_\mathrm{h}}{\partial S}\right) = -4\pi G\left(\rho + p\right)~~,
 \label{HE-4}
\end{eqnarray}
where we have used $S = \frac{\pi}{GH^2}$ is the Bekenstein-Hawking entropy and $\dot{S} = -\frac{2\pi}{G}\left(\frac{\dot{H}}{H^3}\right)$. The above equation acts as the second Friedmann equation in the context of horizon cosmology where the entropy of the apparent horizon is given by $S_\mathrm{h}$. Clearly for $S_\mathrm{h} = S$, i.e. when the entropy of the horizon is given by the Bekenstein-Hawking entropy, Eq.~(\ref{HE-4}) reduces to the usual Friedmann equation for Einstein gravity. Integrating both sides of Eq.~(\ref{HE-4}) by taking the energy conservation of the matter fields into account, yields the following expression:
\begin{eqnarray}
 \int \left(\frac{\partial S_\mathrm{h}}{\partial S}\right) d\left(H^2\right) = \frac{8\pi G}{3}\rho + \frac{\Lambda}{3}~~,
 \label{HE-5}
\end{eqnarray}
where $\Lambda$ is the constant of integration (also known as the cosmological constant), and the integration can be performed once we consider a specific form of the horizon entropy in terms of the Bekenstein-Hawking entropy variable (i.e. $S_\mathrm{h} = S_\mathrm{h}(S)$). Eq.~(\ref{HE-5}) acts as the first Friedmann equation in the horizon cosmology for a general form of the horizon entropy, and once again, it reduces to the usual Friedmann equation for $S_\mathrm{h} = S$. Thus the entropic cosmology with the Bekenstein-Hawking horizon entropy is similar to that of in case of Einstein gravity, otherwise, some other form of the horizon entropy will result to a modified Friedmann equations. For instance, in the case of the Tsallis entropy where $S_\mathrm{h} = S^{\delta}$ (here $\delta$ is a parameter and known as the Tsallis exponent), Eq.~(\ref{HE-5}) and Eq.~(\ref{HE-4}) become,
\begin{eqnarray}
 H^2\left(\frac{\delta}{2-\delta}\right)\left(\frac{\pi}{GH^2}\right)^{\delta-1} = \frac{8\pi G}{3}\rho + \frac{\Lambda}{3}~~,
 \label{HE-T-1}
\end{eqnarray}
and
\begin{eqnarray}
 \delta\left(\frac{\pi}{GH^2}\right)^{\delta-1}\dot{H} = -4\pi G\left(\rho + p\right)~~,
 \label{HE-T-2}
\end{eqnarray}
respectively. The corresponding field equations for other forms of horizon entropies can be similarly obtained, which will be evaluated in following sections. Here it deserves mentioning that in order to derive the Friedmann Eqs.~(\ref{HE-4}) and (\ref{HE-5}), we have used only the first law of thermodynamics of the apparent horizon. However in the context of horizon thermodynamics, a consistent cosmology also demands the validity of second law of thermodynamics, i.e. the change of total entropy (which is the sum of the horizon entropy and the entropy of the matter fields) with cosmic time should be positive. In this regard, beside the thermodynamics of the apparent horizon governed by Eq.~(\ref{law-1}), we also need to consider the thermodynamics of the matter fields.

\section{Thermodynamics of the matter fields inside of the horizon}\label{SecIII}

\begin{figure}[!h]
\begin{center}
\centering
\includegraphics[width=2.5in,height=2.5in]{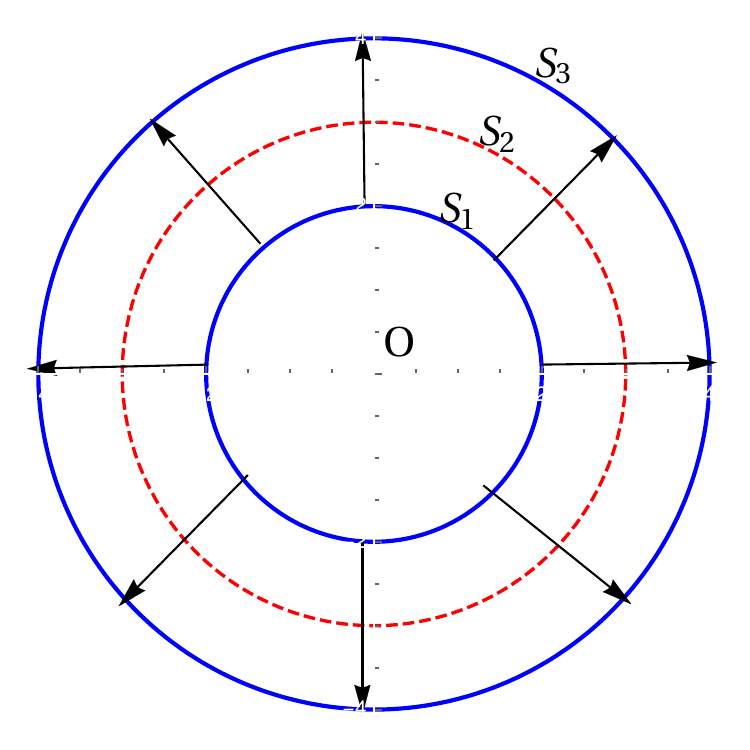}
\caption{Comparison between the formation of apparent horizon and the comoving expansion of the universe, in order to calculate $\frac{dN}{dt}$. The detailed explanation of the figure is given below.}
 \label{plot-2}
\end{center}
\end{figure}

The matter fields inside of the apparent horizon obey the following thermodynamic law:
\begin{eqnarray}
T_\mathrm{m}dS_\mathrm{m} = d\left(\rho V\right) + pdV - \mu dN~~,
\label{ME-1}
\end{eqnarray}
where $T_\mathrm{m}$ and $S_\mathrm{m}$ represent the temperature and the entropy of the matter fields respectively; note that $T_\mathrm{m}$, in general, is different than the horizon temperature (see Sec.~[\ref{sec-total-change}] for the details). As we will discuss later that the matter fields have a flux through the apparent horizon, and moreover, the flux is either outward or inward depending on the background cosmic evolution of the universe. Owing to the presence of such flux, the matter fields obey the thermodynamic law (\ref{ME-1}) applicable for an open system where $\mu$ symbolizes the chemical potential and $dN$ represents the change of matter particles within the horizon in time $dt$. Therefore the effective work done by the matter fields is given by: $dW_\mathrm{m} = pdV - \mu dN$. Eq.~(\ref{ME-1}) immediately leads to,
\begin{eqnarray}
 T_\mathrm{m}\dot{S}_\mathrm{m} = \dot{\rho}V + \left(\rho + p\right)\dot{V} - \mu\dot{N}~~,
 \label{ME-2}
\end{eqnarray}
which, due to $V = \frac{4\pi}{3H^3}$, takes the following form,
\begin{eqnarray}
 T_\mathrm{m}\dot{S}_\mathrm{m} = -\frac{4\pi}{H^2}\left(\rho + p\right)\left\{1 + \frac{\dot{H}}{H^2}\right\} - \mu\dot{N}~~.
 \label{ME-3}
\end{eqnarray}
For a better understanding of $\dot{N}$, we need to understand that the comoving expansion speed of the universe differs from the speed of the formation of apparent horizon. In particular, the comoving speed of the universe at a physical distance $d$ from an observer is given by: $v_\mathrm{c} = Hd$, while the speed of the formation of the apparent horizon comes as $v_\mathrm{h} = -\dot{H}/H^2$. Therefore $v_\mathrm{c} = 1$ at the apparent horizon (i.e. at $d = 1/H$). Thus $v_\mathrm{c} > v_\mathrm{h}$ occurs for an accelerating universe when $-\frac{\dot{H}}{H^2} < 1$; while for a decelerating universe, when $-\frac{\dot{H}}{H^2} > 1$, the comoving expansion of the universe remains less than the speed of the formation of the apparent horizon (i.e. $v_\mathrm{c} < v_\mathrm{h}$). To illustrate this issue, let us focus on Fig.~[\ref{plot-2}] where, for instance, we show the case of $v_\mathrm{c} > v_\mathrm{h}$ (the other case of $v_\mathrm{c} < v_\mathrm{h}$ can be similarly demonstrated).

The concentric spheres in the Fig.~[\ref{plot-2}] denote:
\begin{itemize}
 \item $S_\mathrm{1}$: The visible universe bounded by the apparent horizon at time $t$ with respect to a comoving observer (labeled by O). Therefore the radius of the sphere is given by: $OS_\mathrm{1} = \frac{1}{H(t)}$, and thus the volume of the sphere is $V(t) = 4\pi/(3H^3)$.

 \item $S_\mathrm{2}$: The visible universe bounded by the apparent horizon at time $t+dt$ with respect to a comoving observer (labeled by O). Therefore $OS_\mathrm{2} = \frac{1}{H(t+dt)} = \frac{1}{H} - \frac{\dot{H}}{H^2}dt$ (at the leading order in $dt$), and thus the volume of the sphere is given by,
 \begin{eqnarray}
  V(t+dt) = \frac{4\pi}{3}\left(\frac{1}{H} - \frac{\dot{H}}{H^2}dt\right)^3~~.\nonumber
 \end{eqnarray}
 \item $S_\mathrm{3}$: Due to the difference between $v_\mathrm{c}$ and $v_\mathrm{h}$ (as mentioned earlier), let us consider that the surface $S_\mathrm{1}$ moves from $S_\mathrm{1} \rightarrow S_\mathrm{3}$ (within time $dt$) by the comoving expansion speed. Therefore the volume of the sphere $S_\mathrm{3}$ is given by:
 \begin{eqnarray}
  V_\mathrm{c}(t+dt) = \frac{4\pi}{3}\left(\frac{1}{H} + dt\right)^3~~,\nonumber
 \end{eqnarray}
where we use $v_\mathrm{c}(t) = 1$ at $d=1/H$.
\end{itemize}

In Fig.~[\ref{plot-2}], the visible universe at time $t$ and at $t+dt$ are described by the spheres $S_\mathrm{1}$ and $S_\mathrm{2}$ respectively. However, due to $v_\mathrm{c} \neq v_\mathrm{h}$, the amount of matter fields within $S_\mathrm{1}$ at time $t$ is not equal to that of within $S_\mathrm{2}$ at $t+dt$. This indicate that there exists a flux of the matter fields through the horizon. In order to calculate this, we determine,
\begin{eqnarray}
 V_\mathrm{c}(t+dt) - V(t+dt)&=&\frac{4\pi}{3}\left(\frac{1}{H} - \frac{\dot{H}}{H^2}dt\right)^3 - \frac{4\pi}{3}\left(\frac{1}{H} + dt\right)^3\nonumber\\
 &=&\frac{4\pi}{H^2}\left(1 - \epsilon\right)dt~~,\nonumber
\end{eqnarray}
within the first order in $dt$, or equivalently, we have
\begin{eqnarray}
 \frac{d}{dt}\left[V_\mathrm{c}(t+dt) - V(t+dt)\right] = \frac{4\pi}{H^2}\left(1 - \epsilon\right)~~,
 \label{ME-4}
\end{eqnarray}
where $\epsilon = -\dot{H}/H^2$ (generally known as slow roll parameter). The amount of matter fields enclosed within $V_\mathrm{c}(t+dt)$ and $V(t+dt)$ will eventually lead to the flux of the same through the horizon at time $dt$. Considering the energy per particle to be $u$, we can write
\begin{eqnarray}
 \frac{dN}{dt} = -\frac{1}{u}\frac{d}{dt}\left[V_\mathrm{c}(t+dt) - V(t+dt)\right]~~,
 \label{ME-5}
\end{eqnarray}
where the negative sign indicates that the particle number inside the horizon decreases (with time) when $V_\mathrm{c}(t+dt) > V(t+dt)$ (due to $v_\mathrm{c} > v_\mathrm{h}$ used in the Fig.~[\ref{plot-2}], we get $V_\mathrm{c}(t+dt) > V(t+dt)$. However for a decelerating universe, when $v_\mathrm{c} < v_\mathrm{h}$, one will obtain $V_\mathrm{c}(t+dt) < V(t+dt)$). Using Eq.~(\ref{ME-4}), we immediately get,
\begin{eqnarray}
 \frac{dN}{dt} = \frac{1}{u}\left(\frac{4\pi}{H^2}\right)\left(\epsilon - 1\right)~~.
 \label{ME-5a}
\end{eqnarray}
Eq.~(\ref{ME-5a}) argues that the rate of change of the particle number inside of the horizon in turn depends whether the parameter $\epsilon$ is larger or less than unity. Two different cases appear in this regard --- (a) during the accelerated expansion of the universe when $\epsilon< 1$ (for instance, during the inflation), $\dot{N}$ comes to be negative from Eq.~(\ref{ME-5a}), or equivalently, the matter fields have an outward flux through the horizon, while (b) $\dot{N}$ becomes positive during the decelerated expansion when $\epsilon > 1$ (i.e. during the reheating, and radiation era). With the chemical potential: $\mu \equiv \frac{\partial}{\partial N}\left(\mathrm{total~energy}\right) = u$, we have the following expression from Eq.~(\ref{ME-5a}),
\begin{eqnarray}
 \mu \dot{N} = -\frac{4\pi \rho}{H^2}\left(1-\epsilon\right)~~.
 \label{ME-6}
\end{eqnarray}
Plugging this into Eq.~(\ref{ME-1}) yields,
\begin{eqnarray}
 T_\mathrm{m}\dot{S}_\mathrm{m} = -\frac{4\pi}{H^2}\left(\rho + p\right)\left\{1 + \frac{\dot{H}}{H^2}\right\} + \frac{4\pi \rho}{H^2}\left(1-\epsilon\right)~~,
 \label{ME-7}
\end{eqnarray}
which is the final expression of $\dot{S}_\mathrm{m}$, and we will use this at some later stage.

\section{Change of total entropy with cosmic expansion}\label{sec-total-change}

We start this section by defining the total entropy of the visible universe (bounded by the apparent horizon) as the sum of horizon entropy and the entropy of the matter fields inside of the horizon, in particular,
\begin{eqnarray}
 S_\mathrm{tot} = S_\mathrm{h} + S_\mathrm{m}~~.
 \label{TE-0}
\end{eqnarray}
According to the second law of thermodynamics of the apparent horizon, the change of $S_\mathrm{tot}$ needs to be positive with cosmic expansion of the universe, i.e.
\begin{eqnarray}
 \frac{dS_\mathrm{tot}}{dt} > 0~~~\Longrightarrow~~~\frac{dS_\mathrm{h}}{dt} + \frac{dS_\mathrm{m}}{dt} > 0~~.
 \label{TE--1}
\end{eqnarray}
In the previous sections, Eq.~(\ref{HE-3}) and Eq.~(\ref{ME-7}) provide the change of horizon entropy as well the change of matter fields' entropy (with respect to the cosmic time). Having obtained these, we immediately determine the change of total entropy as,
\begin{eqnarray}
 \dot{S}_\mathrm{h} + \dot{S}_\mathrm{m} = \frac{8\pi}{H^3}\left(\rho + p\right) + \frac{1}{T_\mathrm{m}}\left\{-\frac{4\pi}{H^2}\left(\rho + p\right)\left(1 + \frac{\dot{H}}{H^2}\right) + \frac{4\pi \rho}{H^2}\left(1-\epsilon\right)\right\}~~.
 \label{TE-1}
\end{eqnarray}
Consequently we get,
\begin{eqnarray}
 T_\mathrm{h}\frac{dS_\mathrm{h}}{dt} + T_\mathrm{m}\frac{dS_\mathrm{m}}{dt} = -2\pi\left(\rho + p\right)\left(\frac{\dot{H}}{H^4}\right) + \frac{4\pi \rho}{H^2}\left(1-\epsilon\right)~~,
 \label{TE-2}
\end{eqnarray}
where $T_\mathrm{h}$ and $T_\mathrm{m}$ represent the temperature of the apparent horizon and of the matter field respectively.

The above expression depicts that for a de-Sitter (dS) universe, when $\epsilon = 0$ or $H=\mathrm{constant}$, $T_\mathrm{h}\dot{S}_\mathrm{h} + T_\mathrm{m}\dot{S}_\mathrm{m}$ comes as a positive quantity. Actually the horizon entropy in a dS universe remains constant (with time) --- this is because the apparent horizon becomes static in a dS universe and thus the entropy of the horizon remains constant with time, or equivalently, this can be understood directly from Eq.~(\ref{HE-3}) as $\rho + p = 0$ for $\epsilon = 0$. However the entropy of matter fields inside the horizon changes with the cosmic expansion even in a dS universe. This is due to the fact that $v_\mathrm{c} > v_\mathrm{h}$ in the dS expansion (where $v_\mathrm{h} = 0$ for the dS case, see the discussion about $v_\mathrm{c}$ and $v_\mathrm{h}$ after Eq.~(\ref{ME-3})), and consequently, there exits an outward flux of the matter field through the apparent horizon. Therefore the matter field in a dS universe, which is associated with cosmological constant, follows $\dot{\rho} = \dot{V} = 0$ and $\dot{N} < 0$ (the negative $\dot{N}$ represents the outward flux of the matter field from inside the horizon). Here it deserves mentioning that due to the nature of cosmological constant, the total energy of the matter field inside the horizon (i.e. $E = \rho V$) remains constant despite the existence of the outward flux through the horizon. As a result, the entropy of the matter field in the dS scenario increases with time according to the thermodynamics of the matter field.

Coming back to Eq.~(\ref{TE-2}), it shows that $T_\mathrm{h}\dot{S}_\mathrm{h} + T_\mathrm{m}\dot{S}_\mathrm{m}$ depends on the energy density and the pressure of the matter fields which in turn are controlled by the specific model under consideration. However in order to examine the constraints on entropic parameters in a model independent way direct from the second law of horizon thermodynamics, we need to eliminate $\rho$ and $p$ from the above expression by using the Friedmann Eqs.~(\ref{HE-4}) and (\ref{HE-5}). As a result, we obtain,
\begin{eqnarray}
 T_\mathrm{h}\frac{dS_\mathrm{h}}{dt} + T_\mathrm{m}\frac{dS_\mathrm{m}}{dt} = \frac{\epsilon^2}{2G}\left(\frac{\partial S_\mathrm{h}}{\partial S}\right)
 - \frac{3(\epsilon - 1)}{2G}\frac{1}{H^2}~\int \left(\frac{\partial S_\mathrm{h}}{\partial S}\right) d\left(H^2\right)~~,
 \label{TE-3}
\end{eqnarray}
with recall that $\epsilon = -\dot{H}/H^2$.

Based on \cite{Mimoso:2016jwg}, we may consider that the temperature of the matter fields inside of the horizon coincides with the temperature of the latter except during the radiation era. In particular,
\begin{eqnarray}
 T_\mathrm{h}&\neq&T_\mathrm{m}~~~~~~~~\mathrm{during~radiation~era}~~,\nonumber\\
 T_\mathrm{h}&=&T_\mathrm{m}~~~~~~~~~\mathrm{otherwise}~~.
 \label{TE-3a0}
\end{eqnarray}
The fact that $T_\mathrm{h} \neq T_\mathrm{m}$ in the radiation era is also expected as the horizon temperature vanishes during the same, while it is well known that the temperature of radiation fluid goes by $T_\mathrm{m} \propto a^{-1}$ and hence is non-zero. Therefore other than the radiation era, the second law of thermodynamics of apparent horizon can be equivalently written by,
\begin{eqnarray}
 T_\mathrm{h}\frac{dS_\mathrm{h}}{dt} + T_\mathrm{m}\frac{dS_\mathrm{m}}{dt} > 0~~,
 \label{TE-3a}
\end{eqnarray}
as $T_\mathrm{h} = T_\mathrm{m} > 0$. As a whole, we will use Eq.~(\ref{TE-3}) to examine the validation of the second law of horizon thermodynamics from inflation to reheating era; while such examination during the radiation era needs to be done separately from Eq.~(\ref{TE-3}) because of $T_\mathrm{h} = 0$ and the thermodynamic law (\ref{law-1}) (by using which, Eq.(\ref{TE-3}) is derived) identically vanishes from both sides in the radiation period.

Eq.~(\ref{TE-3}) demonstrates that $T_\mathrm{h}\dot{S}_\mathrm{h} + T_\mathrm{m}\dot{S}_\mathrm{m}$ depends on the form of the horizon entropy as well as on the evolution of the Hubble parameter through $\epsilon = -\dot{H}/H^2$. In the next few subsections, we will consider different forms of the horizon entropy like the Tsallis entropy, the R\'{e}nyi entropy, the Kaniadakis entropy, or even the 4-parameter generalized entropy; and examine the appropriate conditions in order to validate the second law of horizon thermodynamics. Moreover, for each horizon entropy, we will further consider different cosmological epochs of the universe (due to presence of the parameter $\epsilon$ in the rhs of Eq.~(\ref{TE-3})). In this regard, we will particularly concentrate on the following evolutionary stages of the universe: $\mathrm{inflation}~\rightarrow~\mathrm{reheating}~\rightarrow~\mathrm{radiation~era}$ respectively. Thereby the early stage of the universe is described by a de-Sitter (or a quasi de-Sitter) inflation when the Hubble parameter remains almost constant (or equivalently, $\epsilon \simeq 0$). After the inflation ends, the universe enters to the reheating era, during which, the matter energy density decays to relativistic particles with a certain decay width generally considered to be constant, (in the same spirit of \cite{Dai:2014jja,Cook:2015vqa}). During the reheating evolution of the universe, the Hubble parameter is generally parametrized by a power law form of the scale factor i.e. $H(a) \propto a^{-\frac{3}{2}(1+\omega_\mathrm{0})}$ (with $a$ being the scale factor of the universe and $\omega_\mathrm{0}$ is a constant). Here $\omega_\mathrm{0}$, defined by $\omega_\mathrm{0} = -1-2\dot{H}/(3H^2)$, is the EoS parameter of the reheating era and thus related to $\epsilon$ by: $\epsilon = \frac{3}{2}(1+\omega_\mathrm{0})$. Moreover the $\omega_\mathrm{0}$ generally lies between $0 \leq \omega \leq 1$ depending on the background dynamics of the same. Based on the above arguments, we may write the Hubble parameter during inflation and during reheating as:
\begin{itemize}
\item During the inflation: $H=H_\mathrm{I}$ (constant).

 \item During the reheating era: $H(a) = H_\mathrm{f}\left(\frac{a}{a_\mathrm{f}}\right)^{-\frac{3}{2}(1+\omega_\mathrm{0})}$; where $\omega_\mathrm{0}$ is the reheating EoS parameter and $0 \leq \omega_\mathrm{0} \leq 1$.
 \begin{eqnarray}
  \label{N-1}
 \end{eqnarray}
\end{itemize}
Here $H_\mathrm{I}$ is the inflationary energy scale; the suffix 'f' with some quantity denotes the same at the end of inflation, for instance, $a_\mathrm{f}$ is the scale factor at the end of inflation. Clearly the Hubble parameter $H = H(a)$ written in the above fashion is continuous at the junction between two stages. We like to mention that the evolution of the Hubble parameter is governed by Eq.~(\ref{HE-4}) and Eq.~(\ref{HE-5}) for a given form of entropy of the apparent horizon, from which, one can reconstruct $\rho = \rho(a)$ and $p = p(a)$ at different cosmic era by using the corresponding $H = H(a)$ from Eq.~(\ref{N-1}).

\subsection*{\underline{Tsallis entropy}}

For the systems with long range interactions where the Boltzmann-Gibbs entropy is not applied, one needs to introduce the Tsallis entropy which is given by $S_\mathrm{h} \equiv S_\mathrm{T} = S^{\delta}$ (where the suffix 'T' stands for Tsallis entropy and $S = \frac{\pi}{GH^2}$ is the Bekenstein-Hawking entropy), the cosmological field equations are given by Eq.~(\ref{HE-T-1}) and Eq.~(\ref{HE-T-2}) respectively. Owing to the Tsallis entropy, the integral present in the last term of Eq.~(\ref{TE-3}) can be determined as follows:
\begin{eqnarray}
 \frac{1}{H^2}\int \left(\frac{\partial S_\mathrm{T}}{\partial S}\right)d\left(H^2\right) = \frac{\delta}{\left(2-\delta\right)}\left(\frac{\pi}{GH^2}\right)^{\delta - 1}~~.
 \label{T-3}
\end{eqnarray}

Plugging the above expression into Eq.~(\ref{TE-3}), and by using $\frac{\partial S_\mathrm{T}}{\partial S} = \delta\left(\frac{\pi}{GH^2}\right)^{\delta-1}$, yield the change of total entropy, in particular,
\begin{eqnarray}
 T_\mathrm{h}\left(\frac{dS_\mathrm{T}}{dt}\right) + T_\mathrm{m}\left(\frac{dS_\mathrm{m}}{dt}\right) = \left(\frac{\delta}{2G}\right)\left(\frac{\pi}{GH^2}\right)^{\delta - 1}\left\{\epsilon^2 - \frac{3\left(\epsilon - 1\right)}{\left(2 - \delta\right)}\right\}~~.
 \label{T-4}
\end{eqnarray}
Therefore in the case of Tsallis entropy, the quantity $T_\mathrm{h}\dot{S}_\mathrm{T} + T_\mathrm{m}\dot{S}_\mathrm{m}$ takes the above form which needs to be positive according to the second law of thermodynamics of the apparent horizon. As mentioned after Eq.~(\ref{TE-3}) that due to the dependence of $\epsilon$, we will examine the conditions for the positivity of the rhs of Eq.~(\ref{T-4}) by considering different cosmological epochs of the universe from the inflation to the radiation dominated era.

\begin{enumerate}
 \item {\underline{During inflation}}:Here $\epsilon \simeq 0$ (which is well aproximated during inflation in the present context), or equivalently, $H = H_\mathrm{I}$ (constant). As a result, Eq.~(\ref{T-4}) leads to the following expression:
 \begin{eqnarray}
  T_\mathrm{h}\left(\frac{dS_\mathrm{T}}{dt}\right) + T_\mathrm{m}\left(\frac{dS_\mathrm{m}}{dt}\right) = \frac{3}{(2-\delta)}\left(\frac{\delta}{2G}\right)\left(\frac{\pi}{GH^2}\right)^{\delta - 1}~~.
 \label{T-5}
 \end{eqnarray}
 Therefore $T_\mathrm{h}\dot{S}_\mathrm{T} + T_\mathrm{m}\dot{S}_\mathrm{m} > 0$ during inflation requires the constraint on the Tsallis exponent as,
 \begin{eqnarray}
  0 < \delta < 2~~.
  \label{T-6}
 \end{eqnarray}

 \item {\underline{During reheating stage}}: Recall that the EoS parameter during the reheating stage is symbolized by $\omega_\mathrm{0}$ which is related to $\epsilon$ by,
 \begin{eqnarray}
  \epsilon = \frac{3}{2}\left(1+\omega_\mathrm{0}\right)~~.
  \label{T-7}
 \end{eqnarray}
Due to the above form of $\epsilon$, Eq.~(\ref{T-4}) takes the following form:
\begin{eqnarray}
 T_\mathrm{h}\left(\frac{dS_\mathrm{T}}{dt}\right) + T_\mathrm{m}\left(\frac{dS_\mathrm{m}}{dt}\right) = \frac{9}{4}\frac{(1+\omega_\mathrm{0})^2}{(2-\delta)}\left(\frac{\delta}{2G}\right)\left(\frac{\pi}{GH^2}\right)^{\delta - 1}\left\{\frac{2(2+3\omega_\mathrm{0}+3\omega_\mathrm{0}^2)}{3(1+\omega_\mathrm{0})^2} - \delta\right\}~~.
 \label{T-8}
\end{eqnarray}
Consequently $T_\mathrm{h}\dot{S}_\mathrm{T} + T_\mathrm{m}\dot{S}_\mathrm{m} > 0$ during the reheating stage demands the following constraints on the Tsallis exponent as,
\begin{eqnarray}
 0 < \delta < \frac{2(2+3\omega_\mathrm{0}+3\omega_\mathrm{0}^2)}{3(1+\omega_\mathrm{0})^2}~~.
 \label{T-9}
\end{eqnarray}
The fact that the reheating EoS parameter generally lies within $\omega_\mathrm{0} = [0,1]$ allows the quantity $\frac{2(2+3\omega_\mathrm{0}+3\omega_\mathrm{0}^2)}{3(1+\omega_\mathrm{0})^2}$ to have a minimum given by,
\begin{eqnarray}
 \mathrm{Min}~\left(\frac{2(2+3\omega_\mathrm{0}+3\omega_\mathrm{0}^2)}{3(1+\omega_\mathrm{0})^2}\right) = \frac{5}{4}~~,
 \label{T-10}
\end{eqnarray}
and thus Eq.~(\ref{T-9}) is immediately written as,
\begin{eqnarray}
 0 < \delta < \frac{5}{4}~~.
 \label{T-11}
\end{eqnarray}

\item {\underline{During radiation era}}: According to the discussion after Eq.~(\ref{TE-3a}), the second law of horizon thermodynamics during the radiation era needs to be treated separately from Eq.~(\ref{T-4}). Considering the radiation fluid as an ideal Bose gas having temperature $T_\mathrm{m}$, the entropy of the radiation inside of the apparent horizon is given by,
\begin{eqnarray}
 S_\mathrm{m} \propto V T^3_\mathrm{m} \propto \left(\frac{1}{aH}\right)^3~~,
 \label{T-12}
\end{eqnarray}
where we use $V = \frac{4\pi}{3H^3}$ and $T_\mathrm{m} \propto a^{-1}$. Consequently, the change of $S_\mathrm{m}$ (with respect to the cosmic time) is obtained as,
\begin{eqnarray}
 \dot{S}_\mathrm{m} \propto \frac{3}{a^3H^2}\left(\epsilon - 1\right)~~,
 \label{T-12a}
\end{eqnarray}
with $\epsilon = -\dot{H}/H^2$. Since the universe during the radiation stage undergoes through a decelerated expansion, the parameter $\epsilon$ must be larger than unity. This in turn argues from Eq.~(\ref{T-12a}) that the entropy of the radiation fluid inside of the horizon increases with time, in particular,
\begin{eqnarray}
 \dot{S}_\mathrm{m} > 0~~,
 \label{T-12b}
\end{eqnarray}
during the radiation era. Besides the entropy of the matter fields, we also need to calculate the change of the horizon entropy (which is the Tsallis entropy in the present case). For this purpose, by using $S_\mathrm{T} = S^{\delta}$, we get
\begin{eqnarray}
 \dot{S}_\mathrm{T} = -\frac{2\pi\delta}{G} \left(\frac{\pi}{GH^2}\right)^{\delta - 1}\left(\frac{\dot{H}}{H^3}\right)~~,
 \label{T-12c}
\end{eqnarray}
where we use $S = \pi/(GH^2)$. Owing to the Friedmann equations for Tsallis entropy from Eq.~(\ref{HE-T-1}) and Eq.~(\ref{HE-T-2}), the above equation turns out to be,
\begin{eqnarray}
 \dot{S}_\mathrm{T} = \frac{4\pi}{GH}\left(\frac{\delta}{2-\delta}\right) \left(\frac{\pi}{GH^2}\right)^{\delta - 1}~~,
 \label{T-12d}
\end{eqnarray}
which is positive for $\delta < 2$. Therefore Eq.~(\ref{T-12b}) and Eq.~(\ref{T-12d}) clearly argue that the change of total entropy during the radiation era proves to be positive for
\begin{eqnarray}
 0 < \delta < 2~~.
 \label{T-13}
\end{eqnarray}

\end{enumerate}

As a whole, the constraint on the Tsallis exponent, from inflation to radiation dominated era, comes as:
\begin{itemize}
\item During inflation: $0 < \delta < 2$ (from Eq.~(\ref{T-6})).

\item During reheating era: $0 < \delta < \frac{5}{4}$ (from Eq.~(\ref{T-11})).

\item During radiation era: $0  < \delta < 2$ (from Eq.~(\ref{T-13})).
\end{itemize}
Due to the reason that $\delta$ remains constant with the cosmic expansion of the universe, all the above constraints on $\delta$ during different cosmic era get simultaneously fulfilled if it follows
\begin{eqnarray}
 0 < \delta < \mathrm{Min}~\left[2, \frac{5}{4}, 2\right]~~,
 \label{T-16}
\end{eqnarray}
or equivalently,
\begin{eqnarray}
 0 < \delta < \frac{5}{4}~~.
 \label{T-17}
\end{eqnarray}
Therefore in the case of Tsallis entropy, the second law of thermodynamics of apparent horizon is ensured during the entire cosmic evolution of the universe (i.e from $\mathrm{inflation}~\rightarrow~\mathrm{reheating}~\rightarrow~\mathrm{radiation~era}$) if the Tsallis exponent lies within the range given by Eq.~(\ref{T-17}). Here it may be noted that such range of $\delta$ also covers the case of the Bekenstein-Hawking entropy where $\delta = 1$.

\subsection*{\underline{R\'{e}nyi entropy}}

In the case of R\'{e}nyi entropy for the apparent horizon, given by,
\begin{eqnarray}
 S_\mathrm{h} \equiv S_\mathrm{R} = \frac{1}{\alpha}\ln{\left(1 + \alpha S\right)}~~,
 \label{R-0}
\end{eqnarray}
where $\alpha$ is a constant (known as the R\'{e}nyi exponent) and $S = \pi/(GH^2)$ is the Bekenstein-Hawking entropy; the Friedmann equations (i.e. Eq.~(\ref{HE-5}) and Eq.~(\ref{HE-4})) become
\begin{eqnarray}
 H^2\left\{1 - \left(\frac{\pi\alpha}{GH^2}\right)\ln{\left(1 + \frac{GH^2}{\pi\alpha}\right)}\right\} = \frac{8\pi G}{3}\rho~~,
 \label{R-1}
\end{eqnarray}
and
\begin{eqnarray}
 \dot{H}\left\{\frac{GH^2/(\pi\alpha)}{1 + GH^2/(\pi\alpha)}\right\} = -4\pi G\left(\rho + p\right)~~,
 \label{R-2}
\end{eqnarray}
respectively (recall that $\rho$ and $p$ represent the energy density and the pressure for normal matter fields inside of the horizon). With the form of the R\'{e}nyi entropy, the integral in Eq.~(\ref{TE-3}) is evaluated as,
\begin{eqnarray}
 \frac{1}{H^2}\int \left(\frac{\partial S_\mathrm{R}}{\partial S}\right)d\left(H^2\right) = 1 - \left(\frac{\pi\alpha}{GH^2}\right)\ln{\left(1 + \frac{GH^2}{\pi\alpha}\right)}~~,
 \label{R-3}
\end{eqnarray}
and by using the above expression into Eq.~(\ref{TE-3}) yields the following form for the change of total entropy (horizon entropy + entropy of matter fields), in particular, we obtain,
\begin{eqnarray}
 T_\mathrm{h}\frac{dS_\mathrm{R}}{dt} + T_\mathrm{m}\frac{dS_\mathrm{m}}{dt} = \frac{\epsilon^2}{2G}\left(\frac{GH^2/(\pi\alpha)}{1 + GH^2/(\pi\alpha)}\right) - \frac{3(\epsilon - 1)}{2G}\left\{1 - \left(\frac{\pi\alpha}{GH^2}\right)\ln{\left(1 + \frac{GH^2}{\pi\alpha}\right)}\right\}.
 \label{R-4}
\end{eqnarray}
Clearly $T_\mathrm{h}\dot{S}_\mathrm{R} + T_\mathrm{m}\dot{S}_\mathrm{m}$ explicitly depends on the Hubble parameter. Thus the condition $T_\mathrm{h}\dot{S}_\mathrm{R} + T_\mathrm{m}\dot{S}_\mathrm{m} > 0$, coming from the second law of thermodynamics of the apparent horizon, needs to be examined for different cosmic era of the universe (see Eq.~(\ref{N-1}) for the Hubble parameter at different era).

\begin{enumerate}
 \item {\underline{During inflation}}: The slow roll parameter takes $\epsilon \simeq 0$ during inflation, and consequently, Eq.~(\ref{R-4}) is given by,
 \begin{eqnarray}
T_\mathrm{h}\left(\frac{dS_\mathrm{R}}{dt}\right) + T_\mathrm{m}\left(\frac{dS_\mathrm{m}}{dt}\right) = \frac{3}{2G}\left\{1 - \frac{\pi\alpha}{GH_\mathrm{I}^2}\ln{\left(1 + \frac{GH_\mathrm{I}^2}{\pi\alpha}\right)}\right\}~~,
\label{R-5}
 \end{eqnarray}
where, recall that $H_\mathrm{I}$ is considered to be the constant Hubble parameter during the inflation (see Eq.~(\ref{N-1})). Now the function (within the curly braket of Eq.~(\ref{R-5})), namely
\begin{eqnarray}
 f(\frac{\pi\alpha}{GH_\mathrm{I}^2}) = 1 - \frac{\pi\alpha}{GH_\mathrm{I}^2}\ln{\left(1 + \frac{GH_\mathrm{I}^2}{\pi\alpha}\right)}~~,
 \nonumber
\end{eqnarray}
is positive valued for $\alpha > 0$, otherwise, the function is either negative valued or becomes undefined (in particular, $f(\frac{\pi\alpha}{GH_\mathrm{I}^2})$ is not defined in the range $\frac{\pi\alpha}{GH_\mathrm{I}^2} = [-1,0]$). Therefore the condition $T_\mathrm{h}\dot{S}_\mathrm{R} + T_\mathrm{m}\dot{S}_\mathrm{m} > 0$ gets satisfied for positive R\'{e}nyi exponent, i.e. for,
\begin{eqnarray}
 \alpha > 0~~,
 \label{R-6}
\end{eqnarray}
during inflation.

\item {\underline{During reheating stage}}: The Hubble parameter during the reheating stage is shown in Eq.(\ref{N-1}) where the reheating EoS parameter $\omega_\mathrm{0}$ is related to $\epsilon$ by $\epsilon = \frac{3}{2}(1+\omega_\mathrm{0})$. Using these into Eq.~(\ref{R-4}) along with a little bit of simplification, we get,
\begin{eqnarray}
 T_\mathrm{h}\left(\frac{dS_\mathrm{R}}{dt}\right)&+&T_\mathrm{m}\left(\frac{dS_\mathrm{m}}{dt}\right) = \frac{9(1+\omega_\mathrm{0})^2}{8G}\left(\frac{GH^2_\mathrm{f}}{\pi\alpha}\right)\left(\frac{a}{a_\mathrm{f}}\right)^{-3(1+\omega_\mathrm{0})}\left\{1 + \frac{GH^2_\mathrm{f}}{\pi\alpha}\left(\frac{a}{a_\mathrm{f}}\right)^{-3(1+\omega_\mathrm{0})}\right\}^{-1}\nonumber\\
 &-&\frac{3(1+3\omega_\mathrm{0})}{4G}\left\{1 - \frac{\pi\alpha}{GH^2_\mathrm{f}}\left(\frac{a}{a_\mathrm{f}}\right)^{3(1+\omega_\mathrm{0})}\ln{\left(1 + \frac{GH^2_\mathrm{f}}{\pi\alpha}\left(\frac{a}{a_\mathrm{f}}\right)^{-3(1+\omega_\mathrm{0})}\right)}\right\}~~.
 \label{R-7}
\end{eqnarray}
Here $a_\mathrm{f}$ represents the scale factor at the end of inflation, and thus the scale factor during the reheating era obeys $a > a_\mathrm{f}$. Therefore the term present within the curly braket in the second line of Eq.~(\ref{R-7}) becomes positive for
\begin{eqnarray}
 \alpha > GH^2_\mathrm{f}/\pi~~,
 \label{R-8}
\end{eqnarray}
which, in turn, ensures the positivity of $T_\mathrm{h}\dot{S}_\mathrm{R} + T_\mathrm{m}\dot{S}_\mathrm{m}$ during the reheating stage. We would like to mention that $H_\mathrm{f}$, is a model dependent quantity that depends particularly on the specific forms of the energy density and the pressure of the matter fields. Actually Eq.~(\ref{R-1}) and Eq.~(\ref{R-2}), along with some specific forms of $\rho$ and $p$, control the evolution of the Hubble parameter in the case of R\'{e}nyi entropy; and thus the $H_\mathrm{f}$ gets fixed by these equations with certain $\rho$ and $p$. For instance, if the matter field is given by a canonical scalar field, then $\rho = \frac{1}{2}\dot{\Phi}^2 + V(\Phi)$ and $p = \frac{1}{2}\dot{\Phi}^2 - V(\Phi)$ (where $\Phi$ is the scalar field and $V(\Phi)$ is its potential): in this case, the scalar field potential controls the Hubble parameter as per Eq.~(\ref{R-1}) and Eq.~(\ref{R-2}), and thus fixes $H_\mathrm{f}$. However in the current work, rather than considering any particular model, our main motive is to find the constraints on entropic exponent(s) in a $model~independent$ way from second law of horizon thermodynamics.

\item {\underline{During radiation era}}: As we have established in Eq.~(\ref{T-12b}) that the radiation fluid inside the apparent horizon exhibits an increasing entropy with the cosmic expansion, $\dot{S}_\mathrm{m} > 0$ during the radiation dominated stage. Moreover the change of horizon entropy (which is the R\'{e}nyi entropy in the present case) is obtained from Eq.~(\ref{R-0}) as follows:
\begin{eqnarray}
 \dot{S}_\mathrm{R} = -\frac{2\pi}{G}\left(\frac{\dot{H}}{H^3\left(1 + \pi\alpha/(GH^2)\right)}\right)~~,\label{R-9}
\end{eqnarray}
which, due to $\dot{H} < 0$ during radiation era, is positive for $\alpha > 0$. Therefore the validity of the second law of horizon thermodynamics, i.e. $\dot{S}_\mathrm{m} + \dot{S}_\mathrm{R} > 0$, results to the following constraint:
\begin{eqnarray}
 \alpha > 0~~.
 \label{R-10}
\end{eqnarray}
\end{enumerate}

As a whole, the constraint on the R\'{e}nyi exponent, from inflation to radiation dominated era followed by a reheating stage, comes as:
\begin{itemize}
\item During inflation: $\alpha > 0$ (from Eq.~(\ref{R-6})).

\item During reheating era: $\alpha > GH^2_\mathrm{f}/\pi$ (from Eq.~(\ref{R-8})).

\item During radiation era: $\alpha > 0$ (from Eq.~(\ref{R-10})).
\end{itemize}
Being $\alpha$ is a constant with the cosmic expansion of the universe, all the above constraints on $\alpha$ during different cosmic era get concomitantly fulfilled if it obeys
\begin{eqnarray}
 \alpha > \mathrm{Max}~\left[0, GH^2_\mathrm{f}/\pi, 0\right]~~,
 \label{R-13}
\end{eqnarray}
which is equivalent to,
\begin{eqnarray}
 \alpha > GH^2_\mathrm{f}/\pi~~.
 \label{R-14}
\end{eqnarray}
The monotonic decreasing behaviour of the Hubble parameter (with the cosmic time) is indeed ensured from Eq.~(\ref{R-2}) as the matter fields obey the null energy condition during the standard Big-Bang cosmology. Therefore in the case of R\'{e}nyi entropy, the second law of thermodynamics of apparent horizon is valid during the entire cosmic evolution of the universe if the R\'{e}nyi exponent lies within the range given by Eq.~(\ref{R-14}).

\subsection*{\underline{Kaniadakis entropy}}

The Kaniadakis entropy function takes the following form:
\begin{eqnarray}
 S_\mathrm{h} \equiv S_\mathrm{K} = \frac{1}{K}\mathrm{sinh}(KS)~~,
 \label{K-0}
\end{eqnarray}
where $K$ is the Kaniadakis exponent, and once again, $S$ symbolizes the Bekenstein-Hawking entropy. Using $\frac{\partial S_\mathrm{K}}{\partial S} = \mathrm{cosh}(KS)$ and $S=\pi/(GH^2)$, the Friedmann equations, i.e. Eq.~(\ref{HE-4}) and Eq.~(\ref{HE-5}), become,
\begin{eqnarray}
 H^2\left\{\mathrm{cosh}\left(\frac{K\pi}{GH^2}\right) - \left(\frac{K\pi}{GH^2}\right)\mathrm{shi}\left(\frac{K\pi}{GH^2}\right)\right\} = \frac{8\pi G}{3}\rho~~,
 \label{K-1}
\end{eqnarray}
and
\begin{eqnarray}
 \dot{H}\mathrm{cosh}\left(\frac{K\pi}{GH^2}\right) = -4\pi G\left(\rho + p\right)~~,
 \label{K-2}
\end{eqnarray}
respectively. Here $\mathrm{shi}(z)$ is the ``Sinh-Integral'' function and is defined by: $\mathrm{shi}(z) = \int_0^{z} dt \mathrm{sinh}(t)/t$. Moreover the integral in Eq.~(\ref{TE-3}), for Kaniadakis entropy, is obtained as,
\begin{eqnarray}
 \frac{1}{H^2}\int \left(\frac{\partial S_\mathrm{K}}{\partial S}\right)d\left(H^2\right) = \mathrm{cosh}\left(\frac{k\pi}{GH^2}\right) - \left(\frac{k\pi}{GH^2}\right)\mathrm{shi}\left(\frac{k\pi}{GH^2}\right)~~,
 \label{K-3}
\end{eqnarray}
using which into Eq.~(\ref{TE-3}), we get,
\begin{eqnarray}
 T_\mathrm{h}\left(\frac{dS_\mathrm{K}}{dt}\right) + T_\mathrm{m}\left(\frac{dS_\mathrm{m}}{dt}\right)&=&\left(\frac{1}{2G}\right)\epsilon^2~\mathrm{cosh}\left(\frac{k\pi}{GH^2}\right)\nonumber\\
 &-&\left(\frac{3}{2G}\right)\left(\epsilon - 1\right)\left\{\mathrm{cosh}\left(\frac{k\pi}{GH^2}\right) - \left(\frac{k\pi}{GH^2}\right)\mathrm{shi}\left(\frac{k\pi}{GH^2}\right)\right\}~~.
 \label{K-4}
\end{eqnarray}
Having obtained Eq.~(\ref{K-4}), we now examine the condition $T_\mathrm{h}\dot{S}_\mathrm{K} + T_\mathrm{m}\dot{S}_\mathrm{m} > 0$, in the case of Kaniadakis entropy, during different cosmic era of the universe.
\begin{enumerate}
 \item {\underline{During inflation}}: Here $H=H_\mathrm{I}$ or $\epsilon=0$ (see Eq.~(\ref{N-1})), and hence Eq.~(\ref{K-4}) is given by,
 \begin{eqnarray}
T_\mathrm{h}\left(\frac{dS_\mathrm{K}}{dt}\right) + T_\mathrm{m}\left(\frac{dS_\mathrm{m}}{dt}\right) = \frac{3}{2G}\left\{\mathrm{cosh}\left(\frac{K\pi}{GH^2_\mathrm{I}}\right) - \left(\frac{K\pi}{GH^2_\mathrm{I}}\right)\mathrm{shi}\left(\frac{K\pi}{GH^2_\mathrm{I}}\right)\right\}~~.
\label{K-5}
 \end{eqnarray}
 It is clear that the function within the curly braket in the rhs of Eq.~(\ref{K-5}) needs to be positive in order to validate the second law of thermodynamics during the inflation. To have a better understand, we give a plot of the function with respect to the variable $\frac{K\pi}{GH^2}$ in Fig.~[\ref{plot-1}] which demonstrates that the function acquires positive values in the range given by: $-1.4 \lesssim \frac{K\pi}{GH^2_\mathrm{I}} \lesssim 1.4$.
\begin{figure}[!h]
\begin{center}
\centering
\includegraphics[width=3.5in,height=2.5in]{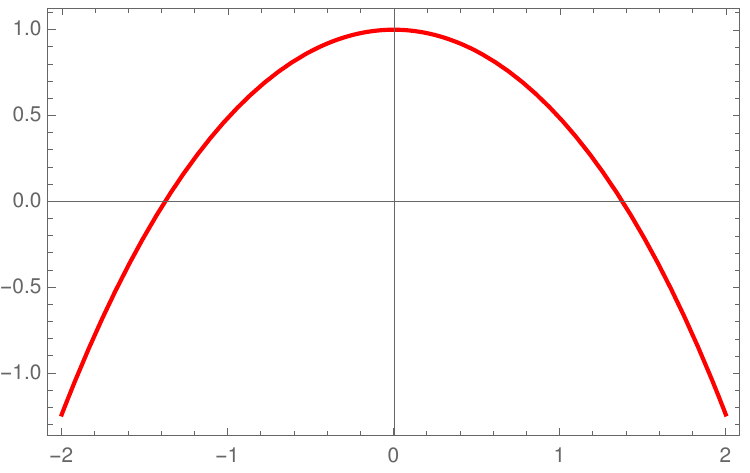}
\caption{The function present within the curly braket in the rhs of Eq.~(\ref{K-5}), namely $F\left(\frac{K\pi}{GH^2_\mathrm{I}}\right) = \mathrm{cosh}\left(\frac{K\pi}{GH^2_\mathrm{I}}\right) - \left(\frac{K\pi}{GH^2_\mathrm{I}}\right)\mathrm{shi}\left(\frac{K\pi}{GH^2_\mathrm{I}}\right)$ (along the vertical axis), with respect to $\frac{K\pi}{GH^2}$ (along the horizontal axis).}
 \label{plot-1}
\end{center}
\end{figure}
Therefor the condition $T_\mathrm{h}\dot{S}_\mathrm{K} + T_\mathrm{m}\dot{S}_\mathrm{m} > 0$ gets satisfied during the inflation for the following range of the Kaniadakis exponent:
\begin{eqnarray}
 -1.4\left(\frac{GH^2_\mathrm{I}}{\pi}\right) \lesssim K \lesssim 1.4\left(\frac{GH^2_\mathrm{I}}{\pi}\right)~~,
 \label{K-6}
\end{eqnarray}
where $H_\mathrm{I}$ is the inflationary Hubble parameter and is a model dependent quantity, as discussed after Eq.~(\ref{R-8}).

\item {\underline{During reheating stage}}: The Hubble parameter evolves according to Eq.(\ref{N-1}), and moreover, the reheating EoS parameter $\omega_\mathrm{0}$ is related to $\epsilon$ by $\epsilon = \frac{3}{2}(1+\omega_\mathrm{0})$. Plugging these into Eq.~(\ref{K-4}) yields the following expression:
\begin{eqnarray}
 T_\mathrm{h}\left(\frac{dS_\mathrm{K}}{dt}\right)&+&T_\mathrm{m}\left(\frac{dS_\mathrm{m}}{dt}\right) = \frac{3(1+3\omega_\mathrm{0}^2)}{8G}~\mathrm{cosh}\left\{\frac{K\pi}{GH^2_\mathrm{f}}\left(\frac{a}{a_\mathrm{f}}\right)^{3(1+\omega_\mathrm{0})}\right\}\nonumber\\
 &+&\frac{3(1+3\omega_\mathrm{0})}{4G}\left(\frac{K\pi}{GH^2_\mathrm{f}}\right)\left(\frac{a}{a_\mathrm{f}}\right)^{3(1+\omega_\mathrm{0})}\mathrm{shi}\left\{\frac{K\pi}{GH^2_\mathrm{f}}\left(\frac{a}{a_\mathrm{f}}\right)^{3(1+\omega_\mathrm{0})}\right\}~~.
 \label{K-7}
\end{eqnarray}
The rhs of Eq.~(\ref{K-7}) contains a ``Cosh'' function and a ``Shi'' function. Owing to the fact that $\mathrm{cosh(z)}$ as well as $z\times\mathrm{shi}(z)$ remain positive for all real $z$, we may argue that the expression in the rhs of Eq.~(\ref{K-7}) is positive for all possible values of Kaniadakis exponent $K$. Therefore in the case of Kaniadakis entropy, the second law of horizon thermodynamics during the reheating stage is ensured for the entire range of $K$.

\item {\underline{During radiation era}}: In the case of Kaniadakis entropy, the change of horizon entropy comes as,
\begin{eqnarray}
 \dot{S}_\mathrm{K} = -\frac{2\pi}{G}\mathrm{cosh}\left(\frac{K\pi}{GH^2}\right)\left(\frac{\dot{H}}{H^3}\right)~~,\nonumber
\end{eqnarray}
and, the change of the entropy for the radiation fluid inside the horizon is given by Eq.~(\ref{T-12a}), i.e.,
\begin{eqnarray}
 \dot{S}_\mathrm{m} \propto \frac{3}{a^3H^2}\left(\epsilon - 1\right)~~.\nonumber
\end{eqnarray}
Therefore the change of the total entropy during the radiation dominated epoch becomes,
\begin{eqnarray}
 \dot{S}_\mathrm{m} + \dot{S}_\mathrm{K} \sim \frac{3}{a^3H^2}\left(\epsilon - 1\right) -\frac{2\pi}{G}\mathrm{cosh}\left(\frac{K\pi}{GH^2}\right)\left(\frac{\dot{H}}{H^3}\right)~~,
 \label{K-8}
\end{eqnarray}
which, due to $\epsilon > 1$, is indeed positive for all possible values of $K$.
\end{enumerate}

As a whole, the constraint on the Kaniadakis exponent, from inflation to reheating, is obtained as:
\begin{itemize}
\item During inflation: $-1.4\left(\frac{GH^2_\mathrm{I}}{\pi}\right) \lesssim K \lesssim 1.4\left(\frac{GH^2_\mathrm{I}}{\pi}\right)$ (from Eq.~(\ref{K-6})).

\item During reheating and radiation era: all possible values of $K$ (see the discussion after Eq.~(\ref{K-7}) and Eq.~(\ref{K-8}) respectively).
\end{itemize}
Due to the fact that $K$ should not vary with the cosmic expansion of the universe as it is a constant, all the above constraints on $K$ during different cosmic era get concomitantly fulfilled if it obeys
\begin{eqnarray}
 -1.4\left(\frac{GH^2_\mathrm{I}}{\pi}\right) \lesssim K \lesssim 1.4\left(\frac{GH^2_\mathrm{I}}{\pi}\right)~~.
 \label{K-10}
\end{eqnarray}
Therefore in the case of Kaniadakis entropy, the second law of thermodynamics of apparent horizon is valid during the entire cosmic evolution of the universe if the Kaniadakis exponent satisfies Eq.~(\ref{K-10}). The inflationary Hubble parameter $H_\mathrm{I}$ can be determined from the Friedmann equations (i.e. Eq.~(\ref{K-1}) and Eq.~(\ref{K-2})) with specific forms of $\rho$ and $p$. However this is not the subject of the present paper as we are interested to determine the constraints on entropic exponent(s) in a model independent way, particularly from the second law of horizon thermodynamics.

\subsection*{\underline{4-parameter generalized entropy}}\label{sec-gen-entropy}

As mentioned in the introduction that recently there has been attempt to generalize the known entropies for the apparent horizon proposed so far (like the Tsallis entropy, the R\'{e}nyi entropy, the Barrow entropy, the Sharma-Mittal entropy, the Kaniadakis entropy and the Loop Quantum gravity entropy). With this motivation, a 6-parameter and a 4-parameter generalized entropy has been proposed in \cite{Nojiri:2022aof} and \cite{Nojiri:2022dkr} respectively, which leads to various forms of horizon entropies (in particular, the Tsallis entropy, the R\'{e}nyi entropy, the Barrow entropy, the Sharma-Mittal entropy, the Kaniadakis entropy and the Loop Quantum gravity entropy) for suitable representation of the entropic parameters. However it has been argued in \cite{Nojiri:2022dkr} that the minimum number of parameters required in a generalized entropy function that can generalize all the aforementioned entropies is equal to four. Thus we will consider the 4-parameter generalized entropy, namely,
\begin{eqnarray}
 S_\mathrm{h} \equiv S_\mathrm{g}\left[\alpha_+,\alpha_-,\beta,\gamma \right] = \frac{1}{\gamma}\left[\left(1 + \frac{\alpha_+}{\beta}~S\right)^{\beta}
 - \left(1 + \frac{\alpha_-}{\beta}~S\right)^{-\beta}\right]~~,
 \label{gen-0}
\end{eqnarray}
in the present context, and will examine its validation under the second law of thermodynamics of the apparent horizon. Here the suffix 'g' in $S_\mathrm{g}$ stands for $generalized$ entropy, and $\alpha_{\pm}$, $\beta$ and $\gamma$ are the corresponding entropic parameters. With the above form of $S_\mathrm{g}$, the corresponding Friedmann equations from Eq.~(\ref{HE-4}) and Eq.~(\ref{HE-5}) take the following form:
\begin{align}
\frac{GH^4\beta}{\pi\gamma}&\,\left[ \frac{1}{\left(2+\beta\right)}\left(\frac{GH^2\beta}{\pi\alpha_-}\right)^{\beta}~
2F_{1}\left(1+\beta, 2+\beta, 3+\beta, -\frac{GH^2\beta}{\pi\alpha_-}\right) \right. \nonumber\\
&\, \left. + \frac{1}{\left(2-\beta\right)}\left(\frac{GH^2\beta}{\pi\alpha_+}
\right)^{-\beta}~2F_{1}\left(1-\beta, 2-\beta, 3-\beta, -\frac{GH^2\beta}{\pi\alpha_+}\right) \right] = \frac{8\pi G\rho}{3} + \frac{\Lambda}{3} \,,
\label{gen-1}
\end{align}
and
\begin{align}
\frac{1}{\gamma}\left[\alpha_{+}\left(1 + \frac{\pi \alpha_+}{\beta GH^2}\right)^{\beta - 1}
+ \alpha_-\left(1 + \frac{\pi \alpha_-}{\beta GH^2}\right)^{-\beta-1}\right]\dot{H} = -4\pi G\left(\rho + p\right)~~,
\label{gen-2}
\end{align}
respectively. Moreover the integral present in Eq.~(\ref{TE-3}), in the case of $S_\mathrm{g}$, comes as,
\begin{align}
 \frac{1}{H^2}\int \left(\frac{\partial S_\mathrm{g}}{\partial S}\right)d\left(H^2\right)&=&\frac{1}{\gamma}\Bigg[\frac{\alpha_{+}}{\left(2-\beta\right)}\left(\frac{GH^2\beta}{\pi\alpha_+}\right)^{1-\beta}~
2F_{1}\left(1-\beta, 2-\beta, 3-\beta, -\frac{GH^2\beta}{\pi\alpha_+}\right)\nonumber\\
&+&\frac{\alpha_-}{\left(2+\beta\right)}\left(\frac{GH^2\beta}{\pi\alpha_-}
\right)^{1+\beta}~2F_{1}\left(1+\beta, 2+\beta, 3+\beta, -\frac{GH^2\beta}{\pi\alpha_-}\right)\Bigg]~~,
 \label{gen-3}
\end{align}
where $2F_{1}[\mathrm{arguments}]$ represents the Hypergeometric function. Plugging the above expression into Eq.~(\ref{TE-3}), we obtain $T_\mathrm{h}\dot{S}_\mathrm{g} + T_\mathrm{m}\dot{S}_\mathrm{m}$ and is given by,
\begin{eqnarray}
 T_\mathrm{h}\left(\frac{dS_\mathrm{g}}{dt}\right)&+&T_\mathrm{m}\left(\frac{dS_\mathrm{m}}{dt}\right) = \left(\frac{1}{2G}\right)\left(\frac{\epsilon^2}{\gamma}\right)\left\{\alpha_+\left(1 + \frac{\pi\alpha_+}{GH^2\beta}\right)^{\beta} + \alpha_-\left(1 + \frac{\pi\alpha_-}{GH^2\beta}\right)^{-\beta}\right\}\nonumber\\
 &-&\left(\frac{3}{2G}\right)\left(\epsilon-1\right)\frac{1}{\gamma}\Bigg[\frac{\alpha_{+}}{\left(2-\beta\right)}\left(\frac{GH^2\beta}{\pi\alpha_+}\right)^{1-\beta}~
2F_{1}\left(1-\beta, 2-\beta, 3-\beta, -\frac{GH^2\beta}{\pi\alpha_+}\right)\nonumber\\
&+&\frac{\alpha_-}{\left(2+\beta\right)}\left(\frac{GH^2\beta}{\pi\alpha_-}
\right)^{1+\beta}~2F_{1}\left(1+\beta, 2+\beta, 3+\beta, -\frac{GH^2\beta}{\pi\alpha_-}\right)\Bigg]~~.
 \label{gen-4}
\end{eqnarray}
Similar to the other cases, here we will also consider various cosmic era (particularly from $\mathrm{inflation}~\rightarrow~\mathrm{reheating}~\rightarrow~\mathrm{radiation~era}$) to investigate the requirement $T_\mathrm{h}\dot{S}_\mathrm{g} + T_\mathrm{m}\dot{S}_\mathrm{m} > 0$ coming from the second law of thermodynamics.

\begin{enumerate}

\item {\underline{During inflation}}: Here $H=H_\mathrm{I}$ or $\epsilon=0$ (see Eq.~(\ref{N-1})), and hence Eq.~(\ref{gen-4}) becomes,
 \begin{eqnarray}
T_\mathrm{h}\frac{dS_\mathrm{g}}{dt} + T_\mathrm{m}\frac{dS_\mathrm{m}}{dt}&=&
\left(\frac{3}{2G}\right)\frac{1}{\gamma}\Bigg[\frac{\alpha_{+}}{\left(2-\beta\right)}\left(\frac{GH^2_\mathrm{I}\beta}{\pi\alpha_+}\right)^{1-\beta}~
2F_{1}\left(1-\beta, 2-\beta, 3-\beta, -\frac{GH^2_\mathrm{I}\beta}{\pi\alpha_+}\right)\nonumber\\
&+&\frac{\alpha_-}{\left(2+\beta\right)}\left(\frac{GH^2_\mathrm{I}\beta}{\pi\alpha_-}
\right)^{1+\beta}~2F_{1}\left(1+\beta, 2+\beta, 3+\beta, -\frac{GH^2_\mathrm{I}\beta}{\pi\alpha_-}\right)\Bigg]~~.
 \label{gen-5}
 \end{eqnarray}
 In order to examine $T_\mathrm{h}\dot{S}_\mathrm{g} + T_\mathrm{m}\dot{S}_\mathrm{m} > 0$ during inflation from Eq.~(\ref{gen-5}), we consider a condition: $\frac{GH^2_\mathrm{I}\beta}{\pi\alpha_{\pm}} < 1$. Such consideration is indeed physical as the inflationary Hubble parameter is generally considered to be less than the Planck scale, for instance, the typical energy scale during inflation is given by $H_\mathrm{I} \sim 10^{-3}/\sqrt{G}$. Owing to $\frac{GH^2_\mathrm{I}\beta}{\pi\alpha_{\pm}} < 1$, the Hypergeometric function in Eq.~(\ref{gen-5}) may be expanded in a Taylor series, and up-to the leading order term, we get:
 \begin{eqnarray}
T_\mathrm{h}\left(\frac{dS_\mathrm{g}}{dt}\right) + T_\mathrm{m}\left(\frac{dS_\mathrm{m}}{dt}\right)&=&
\left(\frac{3}{2G}\right)\frac{1}{\gamma}\Bigg[\frac{\alpha_{+}}{\left(2-\beta\right)}\left(\frac{GH^2_\mathrm{I}\beta}{\pi\alpha_+}\right)^{1-\beta}
\left\{1 - \frac{(1-\beta)(2-\beta)}{(3-\beta)}\left(\frac{GH^2_\mathrm{I}\beta}{\pi\alpha_+}\right)\right\}\nonumber\\
&+&\frac{\alpha_-}{\left(2+\beta\right)}\left(\frac{GH^2_\mathrm{I}\beta}{\pi\alpha_-}
\right)^{1+\beta}\left\{1 - \frac{(1+\beta)(2+\beta)}{(3+\beta)}\left(\frac{GH^2_\mathrm{I}\beta}{\pi\alpha_-}\right)\right\}\Bigg]~~.
 \label{gen-6}
 \end{eqnarray}
Moreover, due to $\frac{GH^2_\mathrm{I}\beta}{\pi\alpha_{\pm}} < 1$, the term containing $\left(\frac{GH^2_\mathrm{I}\beta}{\pi\alpha_{+}}\right)^{1-\beta}$ becomes the dominated one, and thus Eq.~(\ref{gen-6}) may be expressed as,
\begin{eqnarray}
 T_\mathrm{h}\left(\frac{dS_\mathrm{g}}{dt}\right) + T_\mathrm{m}\left(\frac{dS_\mathrm{m}}{dt}\right) \approx
\left(\frac{3}{2G}\right)\frac{\alpha_{+}}{\gamma(2-\beta)}\left(\frac{GH^2_\mathrm{I}\beta}{\pi\alpha_+}\right)^{1-\beta}~~,
 \label{gen-7}
\end{eqnarray}
which is positive for $\gamma > 0$ and $0 < \beta < 2$. Therefore $T_\mathrm{h}\dot{S}_\mathrm{g} + T_\mathrm{m}\dot{S}_\mathrm{m} > 0$ gets fulfilled during the inflation for the following ranges of the entropic parameters:
\begin{eqnarray}
 \frac{\alpha_{\pm}}{\beta} > GH^2_\mathrm{I}/\pi~~~~~~~~~~,~~~~~~~~~~0 < \beta < 2~~~~~~~~\mathrm{and}~~~~~~~\gamma > 0~~.
 \label{gen-8}
\end{eqnarray}

\item {\underline{During reheating stage}}: With $\epsilon = \frac{3}{2}(1+\omega_\mathrm{0})$ during the reheating stage (where $\omega_\mathrm{0}$ is the effective EoS parameyer during the reheating, see Eq.~(\ref{N-1})), and by using $\frac{GH^2\beta}{\pi\alpha_{\pm}} < 1$ (as the Hubble parameter during the cosmic evolution of the universe is well less than the Planck scale), we may write Eq.~(\ref{gen-4}) as,
\begin{eqnarray}
 T_\mathrm{h}\frac{dS_\mathrm{g}}{dt}&+&T_\mathrm{m}\frac{dS_\mathrm{m}}{dt} =
 \left(\frac{1}{2G}\right)\frac{\alpha_{+}}{\gamma}\left(\frac{GH^2\beta}{\pi\alpha_{+}}\right)^{1-\beta}\left[\epsilon^2 - \frac{3(\epsilon-1)}{(2-\beta)}\left\{1 - \frac{(1-\beta)(2-\beta)}{(3-\beta)}\left(\frac{GH^2\beta}{\pi\alpha_{+}}\right)\right\}\right]\nonumber\\
 &+&\left(\frac{1}{2G}\right)\frac{\alpha_{-}}{\gamma}\left(\frac{GH^2\beta}{\pi\alpha_{-}}\right)^{1+\beta}\left[\epsilon^2 - \frac{3(\epsilon-1)}{(2-\beta)}\left\{1 - \frac{(1+\beta)(2+\beta)}{(3+\beta)}\left(\frac{GH^2\beta}{\pi\alpha_{-}}\right)\right\}\right]~~,
 \label{gen-9}
\end{eqnarray}
where $H = H_\mathrm{f}(a/a_\mathrm{f})^{3(1+\omega_\mathrm{0})}$, and once again, we expand the Hypergeometric function of Eq.~(\ref{gen-4}) as a Taylor series (with respect to the variable $\frac{GH^2\beta}{\pi\alpha_{\pm}}$) and retain up-to the leading order term. Moreover, due to $\frac{GH^2_\mathrm{f}\beta}{\pi\alpha_{\pm}} < 1$, the term containing $\left(\frac{GH^2\beta}{\pi\alpha_{+}}\right)^{1-\beta}$ in the above equation contributes the most with respect to the other terms, and thus Eq.~(\ref{gen-9}) becomes,
\begin{eqnarray}
 T_\mathrm{h}\frac{dS_\mathrm{g}}{dt} + T_\mathrm{m}\frac{dS_\mathrm{m}}{dt}&=&
 \left(\frac{9(1+\omega^2_\mathrm{0})}{8G}\right)\frac{\alpha_{+}}{\gamma(2-\beta)}\left(\frac{GH^2_\mathrm{f}\beta}{\pi\alpha_{+}}\left(\frac{a}{a_\mathrm{f}}\right)^{-3(1+\omega_\mathrm{0})}\right)^{1-\beta}\nonumber\\
 &\times&\left[\frac{2(2+3\omega_\mathrm{0} + 3\omega^2_\mathrm{0})}{3(1+\omega_\mathrm{0})^2} - \beta\right]~~,
 \label{gen-10}
\end{eqnarray}
where we use the relation between $\epsilon$ and $\omega_\mathrm{0}$ (as aforementioned). The rhs of Eq.~(\ref{gen-10}), and consequently $T_\mathrm{h}\dot{S}_\mathrm{g} + T_\mathrm{m}\dot{S}_\mathrm{m}$, becomes positive for $\gamma > 0$ and
\begin{eqnarray}
 0 < \beta < \frac{2(2+3\omega_\mathrm{0} + 3\omega^2_\mathrm{0})}{3(1+\omega_\mathrm{0})^2}~~.
 \label{gen-11}
\end{eqnarray}
For $\omega_\mathrm{0}= [0,1]$, we immediately have,
\begin{eqnarray}
 \mathrm{Min}\left[\frac{2(2+3\omega_\mathrm{0} + 3\omega^2_\mathrm{0})}{3(1+\omega_\mathrm{0})^2}\right] = \frac{5}{4}~~,
 \nonumber
\end{eqnarray}
and thus, Eq.~(\ref{gen-11}) is equivalently written as: $0 < \beta < 5/4$. Therefore the validation of the second law of horizon thermodynamics gets ensured if the entropic parameters of $S_\mathrm{g}$ follow:
\begin{eqnarray}
 \frac{\alpha_{\pm}}{\beta} > GH^2_\mathrm{f}/\pi~~~~~~~~~~,~~~~~~~~~~0 < \beta < \frac{5}{4}~~~~~~~~\mathrm{and}~~~~~~~\gamma > 0~~.
 \label{gen-12}
\end{eqnarray}

\item {\underline{During radiation era}}: Using Eq.~(\ref{gen-0}), we determine the change of the 4-parameter generalized entropy with cosmic time, and is given by,
\begin{eqnarray}
 \dot{S}_\mathrm{g} = -\left(\frac{2\pi\dot{H}}{GH^3}\right)\frac{1}{\gamma}\left[\alpha_{+}\left(1 + \frac{\pi \alpha_+}{\beta GH^2}\right)^{\beta - 1}
+ \alpha_-\left(1 + \frac{\pi \alpha_-}{\beta GH^2}\right)^{-\beta-1}\right]~~,
\label{gen-13}
\end{eqnarray}
which, along with Eq.~(\ref{T-12b}), immediately results to the change of the total entropy as,
\begin{eqnarray}
 \dot{S}_\mathrm{m} + \dot{S}_\mathrm{g} \sim \frac{3}{a^3H^2}\left(\epsilon - 1\right) - \left(\frac{2\pi\dot{H}}{GH^3}\right)\frac{1}{\gamma}\left[\alpha_{+}\left(1 + \frac{\pi \alpha_+}{\beta GH^2}\right)^{\beta - 1}
+ \alpha_-\left(1 + \frac{\pi \alpha_-}{\beta GH^2}\right)^{-\beta-1}\right]~~.
\label{gen-13a}
\end{eqnarray}
The above equation clearly indicates that the total entropy during the radiation era (when $\epsilon > 1$) increases with time for the following range of entropic parameters corresponding to $S_\mathrm{g}$:
\begin{eqnarray}
 \alpha_{\pm} > 0~~~~~~~~~~,~~~~~~~~~~\beta > 0~~~~~~~~\mathrm{and}~~~~~~~\gamma > 0~~.
 \label{gen-14}
\end{eqnarray}
\end{enumerate}
As a whole the constraints on the entropic parameters corresponding to the 4-parameter generalized entropy, coming from the validation of the second law of horizon thermodynamics, are given by:
\begin{itemize}
\item During inflation: $\frac{\alpha_{\pm}}{\beta} > GH^2_\mathrm{I}/\pi$; ~~~~$0 < \beta < 2$ and~~~~ $\gamma > 0$ (from Eq.~(\ref{gen-8})).

\item During reheating era: $\frac{\alpha_{\pm}}{\beta} > GH^2_\mathrm{f}/\pi$; ~~~~$0 < \beta < \frac{5}{4}$ and~~~~ $\gamma > 0$ (from Eq.~(\ref{gen-12})).

\item During radiation era: $\frac{\alpha_{\pm}}{\beta} > 0$; ~~~~$\beta > 0$ and~~~~ $\gamma > 0$ (from Eq.~(\ref{gen-14})).
\end{itemize}
Clearly the constraint on $\gamma$ is same during the entire evolution, however $\alpha_{\pm}$ and $\beta$ should follow,
\begin{eqnarray}
 \frac{\alpha_{\pm}}{\beta}&>&\mathrm{Max}\left[GH^2_\mathrm{I}/\pi,~~GH^2_\mathrm{f}/\pi,~~0\right]~~,\nonumber\\
 0 < \beta&<&\mathrm{Min}\left[2, \frac{5}{4}\right]~~,
 \label{gen-17}
\end{eqnarray}
in order to concomitantly satisfy their constraints during all the different cosmic eras. Due to $\dot{H} < 0$ from Eq.~(\ref{gen-2}), the above inequality is immediately written as,
\begin{eqnarray}
 \frac{\alpha_{\pm}}{\beta} > GH^2_\mathrm{I}/\pi~~~~~~~~~\mathrm{and}~~~~~~~~0 < \beta < 5/4~~.
 \label{gen-18}
\end{eqnarray}
Therefore in the context of 4-parameter generalized entropy ($S_\mathrm{g}$), the second law of thermodynamics of apparent horizon is valid during the entire cosmic evolution of the universe (from inflation to radiation dominated era followed by a reheating epoch) if the entropic parameters follow Eq.~(\ref{gen-18}) along with $\gamma > 0$. Here it deserves mentioning that such ranges of $\alpha_{\pm}$, $\beta$ and $\gamma$ in turn make the $S_\mathrm{g}$ (from Eq.~(\ref{gen-0})) as a monotonic increasing function with respect to the Bekenstein-Hawking variable ($S$).

\section{Conclusion}\label{sec-conclusion}

The work investigates the second law of thermodynamics in the context of horizon cosmology, where the universe is described by a spatially flat FLRW metric. Actually in the realm of horizon cosmology, the first law of thermodynamics fixes the cosmological field equations. However a consistent cosmology, besides the first law, also demands the validation of the second law of thermodynamics for the apparent horizon. For this purpose, the present work examines that whether the change of total entropy (i.e. the sum of the entropy of the apparent horizon and the entropy of the matter fields) proves to be positive with the cosmic expansion of the universe. In this regard, the matter fields inside  the horizon show an outward or an inward flux through the apparent horizon depending on whether the universe undergoes  an accelerated or a decelerated expansion respectively. Owing to the presence of such a flux, the matter fields inside the horizon obey the thermodynamics of an open system. It turns out that the change of total entropy (with respect to the cosmic time) depends on the form of horizon entropy as well as on the evolution of the Hubble parameter. Regarding the entropy for the apparent horizon, we consider different forms of the horizon entropy namely the Tsallis entropy, the R\'{e}nyi entropy, the Kaniadakis entropy, or even the 4-parameter generalized entropy; and moreover, for each horizon entropy, we further concentrate on different cosmological epochs of the universe during its evolution history particularly from $\mathrm{inflation}~\rightarrow~\mathrm{reheating}~\rightarrow~\mathrm{radiation~era}$ respectively. Thereby the early stage of the universe is described by a de-Sitter (or a quasi de-Sitter) inflation when the Hubble parameter remains almost constant, and during the reheating stage of the universe, the Hubble parameter is generally parametrized by a power law form of the scale factor i.e. $H(a) \propto a^{-\frac{3}{2}(1+\omega_\mathrm{0})}$ (with $a$ being the scale factor of the universe and $\omega_\mathrm{0}$ is the effective EoS parameter of the reheating era). With such considerations, we determine the appropriate conditions on the respective entropic parameters (for different horizon entropies aforementioned) in order to validate the second law of thermodynamics from inflation to radiation dominated era followed by a reheating stage. Here it deserves mentioning that the constraints on the entropic parameters in turn make the respective entropy as a monotonic increasing function with respect to the Bekenstein-Hawking entropy variable.

In summary, the current work provides model independent constraints on entropic parameters (for different entropy functions of apparent horizon) directly from the second law of horizon thermodynamics during a wide range of cosmic era of the universe.

Finally we would like to mention that in the current work, we do not consider the dark energy epoch of the universe. However the investigation of the second law of horizon thermodynamics during the dark energy era is important from its own right, as it may help to understand the late time acceleration of the universe directly from the second law of horizon thermodynamics. We hope to consider this issue at some future work.

\section*{Acknowledgments}
This work was partially supported by MICINN (Spain), project PID2019-104397GB-I00 and by the program Unidad de Excelencia Maria de Maeztu CEX2020-001058-M, Spain (S.D.O).


\begin{thebibliography}{99}

\bibitem{Bekenstein:1973ur}
J.~D.~Bekenstein,
Phys. Rev. D \textbf{7} (1973), 2333-2346
doi:10.1103/PhysRevD.7.2333


\bibitem{Hawking:1975vcx}
S.~W.~Hawking,
Commun. Math. Phys. \textbf{43} (1975), 199-220
[erratum: Commun. Math. Phys. \textbf{46} (1976), 206]
doi:10.1007/BF02345020

\bibitem{Bardeen:1973gs}
J.~M.~Bardeen, B.~Carter and S.~W.~Hawking,
Commun. Math. Phys. \textbf{31} (1973), 161-170
doi:10.1007/BF01645742

\bibitem{Wald:1999vt}
R.~M.~Wald,
Living Rev. Rel. \textbf{4} (2001), 6
doi:10.12942/lrr-2001-6
[arXiv:gr-qc/9912119 [gr-qc]].

\bibitem{Tsallis:1987eu}
C.~Tsallis,
J. Statist. Phys. \textbf{52} (1988), 479-487
doi:10.1007/BF01016429

\bibitem{Renyi}
A.~R\'{e}nyi, Proceedings of the Fourth Berkeley Symposium on Mathematics, Statistics and Probability, University of California Press (1960), 547-56.

\bibitem{Barrow:2020tzx}
J.~D.~Barrow,
Phys. Lett. B \textbf{808} (2020), 135643
doi:10.1016/j.physletb.2020.135643
[arXiv:2004.09444 [gr-qc]].

\bibitem{SayahianJahromi:2018irq}
A.~Sayahian Jahromi, S.~A.~Moosavi, H.~Moradpour, J.~P.~Morais Gra\c{c}a, I.~P.~Lobo, I.~G.~Salako and A.~Jawad,
Phys. Lett. B \textbf{780} (2018), 21-24
doi:10.1016/j.physletb.2018.02.052
[arXiv:1802.07722 [gr-qc]].

\bibitem{Kaniadakis:2005zk}
G.~Kaniadakis,
Phys. Rev. E \textbf{72} (2005), 036108
doi:10.1103/PhysRevE.72.036108
[arXiv:cond-mat/0507311 [cond-mat]].


\bibitem{Majhi:2017zao}
A.~Majhi,
Phys. Lett. B \textbf{775} (2017), 32-36
doi:10.1016/j.physletb.2017.10.043
[arXiv:1703.09355 [gr-qc]].

\bibitem{Nojiri:2022aof}
S.~Nojiri, S.~D.~Odintsov and V.~Faraoni,
Phys. Rev. D \textbf{105} (2022) no.4, 044042
doi:10.1103/PhysRevD.105.044042
[arXiv:2201.02424 [gr-qc]].

\bibitem{Nojiri:2022dkr}
S.~Nojiri, S.~D.~Odintsov and T.~Paul,
Phys. Lett. B \textbf{831} (2022), 137189
doi:10.1016/j.physletb.2022.137189
[arXiv:2205.08876 [gr-qc]].

\bibitem{Odintsov:2022qnn}
S.~D.~Odintsov and T.~Paul,
Phys. Dark Univ. \textbf{39} (2023), 101159
doi:10.1016/j.dark.2022.101159
[arXiv:2212.05531 [gr-qc]].

\bibitem{Nojiri:2022nmu}
S.~Nojiri, S.~D.~Odintsov and T.~Paul,
Phys. Lett. B \textbf{835} (2022), 137553
doi:10.1016/j.physletb.2022.137553
[arXiv:2211.02822 [gr-qc]].

\bibitem{Odintsov:2023qfj}
S.~D.~Odintsov and T.~Paul,
[arXiv:2301.01013 [gr-qc]].

\bibitem{Odintsov:2023vpj}
S.~D.~Odintsov, S.~D'Onofrio and T.~Paul,
Phys. Dark Univ. \textbf{42} (2023), 101277
doi:10.1016/j.dark.2023.101277
[arXiv:2306.15225 [gr-qc]].

\bibitem{Cai:2005ra}
R.~G.~Cai and S.~P.~Kim,
JHEP {\bf 0502}, 050 (2005)
[arXiv:hep-th/0501055].


\bibitem{Akbar:2006kj}
M.~Akbar and R.~G.~Cai,
Phys.\ Rev.\ D {\bf 75}, 084003 (2007)
[hep-th/0609128].

\bibitem{Cai:2006rs}
R.~G.~Cai and L.~M.~Cao,
Phys.\ Rev.\ D {\bf 75}, 064008 (2007)
[gr-qc/0611071].



\bibitem{Paranjape:2006ca}
A.~Paranjape, S.~Sarkar and T.~Padmanabhan,
Phys.\ Rev.\ D {\bf 74}, 104015 (2006)
[hep-th/0607240].





\bibitem{Jamil:2009eb}
M.~Jamil, E.~N.~Saridakis and M.~R.~Setare,
Phys.\ Rev.\ D {\bf 81}, 023007 (2010)
[arXiv:0910.0822 [hep-th]].

\bibitem{Cai:2009ph}
R.~G.~Cai and N.~Ohta,
Phys.\ Rev.\ D {\bf 81}, 084061 (2010)
[arXiv:0910.2307 [hep-th]].


\bibitem{Jamil:2010di}
M.~Jamil, E.~N.~Saridakis and M.~R.~Setare,
JCAP {\bf 1011}, 032 (2010)
[arXiv:1003.0876 [hep-th]].

\bibitem{Gim:2014nba}
Y.~Gim, W.~Kim and S.~H.~Yi,
JHEP {\bf 1407}, 002 (2014)
[arXiv:1403.4704 [hep-th]].





\bibitem{DAgostino:2019wko}
R.~D'Agostino,
Phys.\ Rev.\ D {\bf 99} (2019) no.10, 103524
[arXiv:1903.03836 [gr-qc]].


\bibitem{Nojiri:2023wzz}
S.~Nojiri, S.~D.~Odintsov, T.~Paul and S.~SenGupta,
Phys. Rev. D \textbf{109} (2024) no.4, 043532
doi:10.1103/PhysRevD.109.043532
[arXiv:2307.05011 [gr-qc]].



\bibitem{Sanchez:2022xfh}
L.~M.~Sanchez and H.~Quevedo,
[arXiv:2208.05729 [gr-qc]].


\bibitem{Cognola:2005de}
G.~Cognola, E.~Elizalde, S.~Nojiri, S.~D.~Odintsov and S.~Zerbini,
JCAP \textbf{02} (2005), 010
doi:10.1088/1475-7516/2005/02/010
[arXiv:hep-th/0501096 [hep-th]].


\bibitem{Nojiri:2021iko}
S.~Nojiri, S.~D.~Odintsov and T.~Paul,
Symmetry \textbf{13} (2021) no.6, 928
doi:10.3390/sym13060928
[arXiv:2105.08438 [gr-qc]].



\bibitem{Witten:1998qj}
E.~Witten,
Adv. Theor. Math. Phys. \textbf{2} (1998), 253-291
doi:10.4310/ATMP.1998.v2.n2.a2
[arXiv:hep-th/9802150 [hep-th]].


\bibitem{Susskind:1998dq}
L.~Susskind and E.~Witten,
[arXiv:hep-th/9805114 [hep-th]].


\bibitem{Fischler:1998st}
W.~Fischler and L.~Susskind,
[arXiv:hep-th/9806039 [hep-th]].




\bibitem{Li:2004rb}
M.~Li,
Phys.\ Lett.\ B {\bf 603} (2004) 1
doi:10.1016/j.physletb.2004.10.014
[hep-th/0403127].


\bibitem{Wang:2016och}
S.~Wang, Y.~Wang and M.~Li,
Phys.\ Rept.\  {\bf 696} (2017) 1
doi:10.1016/j.physrep.2017.06.003
[arXiv:1612.00345 [astro-ph.CO]].

\bibitem{Pavon:2005yx}
D.~Pavon and W.~Zimdahl,
Phys.\ Lett.\ B {\bf 628} (2005) 206
doi:10.1016/j.physletb.2005.08.134
[gr-qc/0505020].

\bibitem{Nojiri:2005pu}
S.~Nojiri and S.~D.~Odintsov,
Gen.\ Rel.\ Grav.\  {\bf 38} (2006) 1285
doi:10.1007/s10714-006-0301-6
[hep-th/0506212].






\bibitem{Malekjani:2012bw}
M.~Malekjani,
Astrophys.\ Space Sci.\  {\bf 347} (2013) 405
doi:10.1007/s10509-013-1522-2
[arXiv:1209.5512 [gr-qc]].


\bibitem{Khurshudyan:2016gmb}
M.~Khurshudyan,
Astrophys. Space Sci. \textbf{361} (2016) no.12, 392
doi:10.1007/s10509-016-2981-z



\bibitem{Gao:2007ep}
C.~Gao, F.~Wu, X.~Chen and Y.~G.~Shen,
Phys.\ Rev.\ D {\bf 79} (2009) 043511
doi:10.1103/PhysRevD.79.043511
[arXiv:0712.1394 [astro-ph]].





\bibitem{Zhang:2005hs}
X.~Zhang and F.~Q.~Wu,
Phys.\ Rev.\ D {\bf 72} (2005) 043524
doi:10.1103/PhysRevD.72.043524
[astro-ph/0506310].

\bibitem{Li:2009bn}
M.~Li, X.~D.~Li, S.~Wang and X.~Zhang,
JCAP {\bf 0906} (2009) 036
doi:10.1088/1475-7516/2009/06/036
[arXiv:0904.0928 [astro-ph.CO]].

\bibitem{Feng:2007wn}
C.~Feng, B.~Wang, Y.~Gong and R.~K.~Su,
JCAP {\bf 0709} (2007) 005
doi:10.1088/1475-7516/2007/09/005
[arXiv:0706.4033 [astro-ph]].

\bibitem{Zhang:2009un}
X.~Zhang,
Phys.\ Rev.\ D {\bf 79} (2009) 103509
doi:10.1103/PhysRevD.79.103509
[arXiv:0901.2262 [astro-ph.CO]].

\bibitem{Lu:2009iv}
J.~Lu, E.~N.~Saridakis, M.~R.~Setare and L.~Xu,
JCAP {\bf 1003} (2010) 031
doi:10.1088/1475-7516/2010/03/031
[arXiv:0912.0923 [astro-ph.CO]].



\bibitem{Nojiri:2017opc}
S.~Nojiri and S.~Odintsov,
Eur. Phys. J. C \textbf{77} (2017) no.8, 528
doi:10.1140/epjc/s10052-017-5097-x
[arXiv:1703.06372 [hep-th]].





\bibitem{Nojiri:2019kkp}
S.~Nojiri, S.~D.~Odintsov and E.~N.~Saridakis,
Phys. Lett. B \textbf{797} (2019), 134829
doi:10.1016/j.physletb.2019.134829
[arXiv:1904.01345 [gr-qc]].



\bibitem{Oliveros:2019rnq}
A.~Oliveros and M.~A.~Acero,
EPL \textbf{128} (2019) no.5, 59001
doi:10.1209/0295-5075/128/59001
[arXiv:1911.04482 [gr-qc]].



\bibitem{Nojiri:2020wmh}
S.~Nojiri, S.~D.~Odintsov, V.~K.~Oikonomou and T.~Paul,
Phys. Rev. D \textbf{102} (2020) no.2, 023540
doi:10.1103/PhysRevD.102.023540
[arXiv:2007.06829 [gr-qc]].







\bibitem{Nojiri:2023bom}
S.~Nojiri, S.~D.~Odintsov and T.~Paul,
Phys. Lett. B \textbf{847} (2023), 138321
doi:10.1016/j.physletb.2023.138321
[arXiv:2311.03848 [gr-qc]].

\bibitem{Gong:2007md}
Y.~Gong and A.~Wang,
Phys. Rev. Lett. \textbf{99} (2007), 211301
doi:10.1103/PhysRevLett.99.211301
[arXiv:0704.0793 [hep-th]].


\bibitem{Padmanabhan:2002sha}
T.~Padmanabhan,
Class. Quant. Grav. \textbf{19} (2002), 5387-5408
doi:10.1088/0264-9381/19/21/306
[arXiv:gr-qc/0204019 [gr-qc]].


\bibitem{Mimoso:2016jwg}
J.~P.~Mimoso and D.~Pav\'on,
Phys. Rev. D \textbf{94} (2016) no.10, 103507
doi:10.1103/PhysRevD.94.103507
[arXiv:1610.07788 [gr-qc]].

\bibitem{Dai:2014jja}
L.~Dai, M.~Kamionkowski and J.~Wang,
Phys. Rev. Lett. \textbf{113} (2014), 041302
doi:10.1103/PhysRevLett.113.041302
[arXiv:1404.6704 [astro-ph.CO]].

\bibitem{Cook:2015vqa}
J.~L.~Cook, E.~Dimastrogiovanni, D.~A.~Easson and L.~M.~Krauss,
JCAP \textbf{04} (2015), 047
doi:10.1088/1475-7516/2015/04/047
[arXiv:1502.04673 [astro-ph.CO]].


\end{thebibliography}
\end{document}